\documentclass[english,prd,superscriptaddress,nofootinbib,preprintnumbers,twocolumn,showpacs]{revtex4}
\usepackage[utf8]{inputenc}
\usepackage[english]{babel}
\usepackage{amsmath}
\usepackage{amsfonts}
\usepackage{amssymb}
\usepackage{epsfig}
\usepackage{graphics,psfrag,rotating}
\usepackage{graphicx}
\usepackage{dcolumn}
\usepackage{bm}
\usepackage{epstopdf}
\usepackage{color}
\usepackage[usenames,dvipsnames,svgnames]{xcolor}
\usepackage[T1]{fontenc}
\usepackage{multirow}
\usepackage{float}

\usepackage{subfigure}

\usepackage{enumitem}
\usepackage[colorlinks=true,
            linkcolor=red,
          urlcolor=gray,
            citecolor=blue]{hyperref}

\def\3nab{\tilde{\nabla}}

\def\be {\begin{equation}}
\def\ee {\end{equation}}
\def\ba {\begin{align}}
\def\ea {\end{align}}

\def\bc {\begin{center}}
\def\ec {\end{center}}
\def\case#1/#2{\frac{#1}{#2}}

\newcommand{\bver}{\begin{verbatim}}
\newcommand{\ever}{\end{verbatim}}

\newcommand{\bea}{\begin{eqnarray}}
\newcommand{\eea}{\end{eqnarray}}
\newcommand{\beaa}{\begin{eqnarray*}}
\newcommand{\eeaa}{\end{eqnarray*}}

\def\case#1/#2{\textstyle\frac{#1}{#2}}

\begin{document}

\title{The realistic models of relativistic stars in $f(R)=R+\alpha R^2$ gravity}

\author{Artyom V. Astashenok\footnote{aastashenok[at]kantiana.ru}}
\affiliation{I. Kant Baltic Federal University, Institute of
Physics, Mathematics and IT, Nevskogo st. 14, 236016 Kaliningrad,
Russia}
\author{Sergei D. Odintsov\footnote{odintsov[at]ieec.uab.es}
} \affiliation{Instituci\`{o} Catalana de Recerca i Estudis
Avan\c{c}ats (ICREA), Barcelona, Spain} \affiliation{Institute of
Space Sciences (IEEC-CSIC) C. Can Magrans s/n, 08193 Barcelona,
Spain}
\author{\'Alvaro de la Cruz-Dombriz\footnote{ alvaro.delacruzdombriz [at] uct.ac.za }}
\affiliation{Cosmology and Gravity Group, Department of Mathematics and Applied Mathematics, University of Cape Town, Rondebosch 7701, Cape Town, South Africa.}



\begin{abstract}
In the context of $f(R)=R+\alpha R^{2}$ gravity, we study the existence of neutron and quark stars for various $\alpha$ with no intermediate approximation in the system of equations.
Analysis shows that for positive $\alpha$
the scalar
curvature does not drop to zero at the star surface (as in General
Relativity) but exponentially decreases with distance. Also
the stellar mass bounded by star
surface decreases when the value $\alpha$ increases. Nonetheless distant
observers would observe a gravitational mass 
due to appearance of  a so-called {\it gravitational sphere} around the star.
The non-zero curvature contribution to the gravitational mass eventually is shown to compensate the stellar mass decrease for growing $\alpha$'s.
We perform our analysis for several equations of state including purely hadronic configurations as well as
hyperons and quark stars. In all cases, we assess that
the relation between the parameter $\alpha$ and the gravitational mass
weakly depends upon the chosen equation of state.
%
Another interesting feature is  the increase of the star radius in comparison with General
Relativity for stars with masses close to maximal,
whereas for intermediate masses $1.4-1.6 M_\odot$ the radius of star depends upon $\alpha$ very weakly.
Also the decrease in the mass bounded by star surface may cause the surface redshift to decrease in $R^2$-gravity when compared to Einsteinian predictions.  This effect is shown to hardly depend upon the observed gravitational mass.
Finally, for negative values of $\alpha$ our analysis shows that outside the
star the scalar curvature has damped oscillations but the contribution
of the {\it gravitational sphere} into the gravitational mass increases
indefinitely with radial distance putting into question the very existence of such relativistic stars.

\end{abstract}


\maketitle

\section{Introduction}

The discovery of the accelerated expansion of the universe \cite{Riess, Perlmutter} has recently led to an intensive study of cosmological applications of extended (also dubbed modified) theories
of gravity which aim to overcome the limitations of the cosmological Concordance $\Lambda$CDM model, made of General Relativity (GR), a cosmological constant $\Lambda$ and Cold (non-baryonic) Dark Matter. Many of these extensions
aim to avoid the usually invoked need of dark components which is required, for instance both in the $\Lambda$CDM model, where dark energy is
nothing but a cosmological constant, and in phantom/quintessence-like models where a scalar field is needed, among others.
Throughout the years, a plethora of modified gravity theories have been shown to be able to preserve the positive results of
Einstein theory and obtain the required cosmological evolution with the most relevant cosmological epochs
\cite{Capozziello2, Odintsov1, Turner}.
Among those attempts, one of the most
successful ones is the so-called $f(R)$ theories where the gravitational
Lagrangian includes powers of the Ricci scalar $R$ encompassed in
an arbitrary function of $R$ ({\it
c.f.}~\cite{Odintsov-3, Capozziello_book, Capozziello4} for extensive reviews
on the subject).

Since in principle the number of viable modified gravities, and
classes of models therein, is sufficiently large, the comparison
between theoretical predictions and different catalogues of
cosmological dataset, extracted from both the cosmic background
and relativistic perturbations (mainly scalar and tensor)
 evolutions, remains mandatory.  Usually,
such comparisons suffer from the so-called degeneracy problem,
meaning that several competitive gravitational theories are able to
explain  the same phenomena
 with roughly speaking the same statistical precision.
Other approaches recently explored have considered
model-independent (cosmographic) techniques in order to
reconstruct the underlying theory capable of causing the observed cosmological
evolution once the Copernican principle is assumed ({\it c.f.} \cite{Cruz-Cosmography,Bamba:2012}
and references therein).

Thus in order to further constrain the number of experimentally
viable theories beyond Einsteinian gravity and obtain more rigid
constraints on the parameters space of every theory, one
possibility consists of studying the consequences of these
proposals at the astrophysical level.
Understandably, these high-energy configurations are the ideal arena to study the behaviour
of extended theories of gravity in strong gravitational field regimes and may complement the picture
as provided by the low-curvature late-time cosmological evolution. First of all, it is interesting to
investigate the possibility of the existence and features of black holes \cite{delaCruzDombriz:2012xy} and relativistic stars,
such as white dwarfs and neutron stars, the latter being the objective of the research in this paper.
Thus for viable classes of $f(R)$ Lagrangians, one needs to obtain the relations between the paradigmatic
parameters of relativistic stars such as masses, radii, momenta of
inertia, quadrupole momenta and Love number, among others. Unfortunately, at the present,
fully precise astrophysical measurements cannot be provided for the aforementioned quantities,
the exception being the mass measurements for neutron stars. 

As extensively studied, once GR is assumed as the correct theory
of gravity, there are certainly upper limits for realistic neutron
star masses, being the theoretical mass limit itself increased
along the years, from $\mathcal{O}(0.6 M_{\odot})$ that
Tolman-Oppenheimer-Volkoff (TOV)~\cite{Oppenheimer:1939ne} found
for a free neutron gas equation of state (EoS),
to $\mathcal{O}(2.2M_{\odot})$ limits when stiffer chiral interactions are allowed~\cite{Dobado:2011gd}.
Notwithstanding, the discovery  of a neutron star with mass $1.974M_{\odot}$ \cite{Demorest:2010bx} by the Shapiro delay method, confirmed some years after by a precise sighting~\cite{Antoniadis:2013pzd} through binary system measurements, have indeed put into question the validity of GR in such high-energy gravitational environment and have been able to rule out exotic EoS proposals. Further observations
have certainly confirmed the existence of neutron stars heavier than the GR naive prediction~\cite{others_masses} as well as double neutron stars and pulsar systems 
also violating the standard GR predictions ({\it c.f.} \cite{Dexheimer:2007mt} for an exhaustive list of the latter) . In that realm,  some attempts have also tried  to reconcile these results with GR, assuming for instance strong magnetic fields \cite{Astashenok-1} or electric charges~\cite{Jing:2015ota} inside the star.

Neutron stars in the context of  $f(R)$ theories were primarily studied in \cite{Kobayashi-Maeda, Upadhye-Hu} and then in abundant bibliography
\cite{Feng:2017hje, Pannia:2016qbj, Wojnar:2016bzk, Arapoglu:2016ozr, Katsuragawa:2015lbl, Fiziev:2015xpa, Hendi:2015pua, Momeni:2015vwa, Zubair:2016kov,Bakirova:2016ffk, Resco:2016upv, Moraes:2015uxq, Sharif:2015jaa, Sotani:2017pfj}
for both  scalar-tensor $f(R)$ theories, as well as
some closely-related theories.
%
Also the possibility of emergence of new types of astrophysical objects in the context of $f(R)$ theories, such as stable stars with large magnetic fields, supermassive stars, among others (see for instance \cite{Laurentis, Laurentis2} for some proposals).

Perturbative solutions of
TOV equations for simple $f(R)$ models
were investigated in \cite{Arapoglu, Alavirad, Astashenok-1}. Obviously, limitations of a perturbative approach lie mainly in the impossibility of
comparing the (unknown) exact solution with the determined perturbative solution. Also one would expect that in strong gravity regimes, non-GR gravitational effects to be dominant and consequently, the GR limit cannot be simply considered as the leading contribution in the solution. Thus, for a careful description one needs to solve
exact higher-order differential equations as we shall do herein.

Another approach in the literature has consisted of resorting to
the equivalence between $f(R)$ theories and Brans-Dicke
scalar-tensor theory in order to study realistic models of neutron
and strange stars \cite{Yazadjiev2014,Doneva2014a,Henttunen}. Within this approach the description in terms of
$f(R)=R+\alpha R^2$ gravity was proposed in \cite{Astashenok2015}
and illustrated for quark stars. There, it was shown that for distant observers the
gravitational mass of star increases with increasing $\alpha$
($\alpha>0$), though the interpretation of this fact depends upon the
frame in which calculations are performed.

Considering $f(R)$ theory directly, one can see that the matching conditions at the edge of
the star, do not impose the scalar curvature to vanish \cite{Clifton:2012ry}, so $R\neq 0$ is in principle expected outside the star. As a
consequence, the spacetime region outside the star also
contributes to the total gravitational mass as perceived by a
distant observer ({\it c.f.} \cite{Resco:2016upv} for a thorough
discussion on this issue). By analogy with scalar-tensor theory of
gravity in which that region outside the star is referred to as
the dilaton sphere (or disphere) \cite{Fiziev}, in the following
we shall denote this domain as {\it gravitational sphere} (or {\it
gravisphere}).

In this paper we shall consider in detail the possibility of
existence and main features of neutron stars and quark (also
dubbed strange) stars in the context of $f(R)=R+\alpha R^2$
gravity. The TOV set of equations for $f(R)$ theories
is presented in the Section \ref{Sec:II}. Then we shall  briefly
discuss the various equations of state (EoS) for nuclear matter in
Sec. \ref{Sec:III}. As mentioned above, for modified gravity
theories, in principle one needs to investigate the solution of
field equations outside the star since it may happen that the
exterior spacetime is not Ricci flat and Schwarzschild is not the
only permitted solution. As we shall show, the obtained exterior
solution crucially depends upon the chosen EoS because the latter fixes the
scalar curvature value at the star surface.
The next two following section will be devoted to obtaining
solutions for various EoS. In Sec. \ref{Sec:IV} we shall show how
for positive $\alpha$ one needs to choose the initial condition
for $R(0)$ to satisfy the asymptotic (Schwarzschild) flatness requirement. The
scalar curvature outside the star thus sharply decreases. It will be
shown that the stellar mass, for a given
central density (i.e., density at the centre of the star)
is smaller than the GR counterpart. Nonetheless the
emergence of the {\it gravitational sphere} around the star,
provides an extra contribution to the gravitational mass as
measured by a distant observer.
As a result the gravitational mass overcomes the GR counterpart for
central density values higher than a given critical central density. 
The value of this critical central density will be shown to 
be determined by both $\alpha$ and the chosen EoS.
This effect takes place for various EoS, including the exotic case of quark stars. It is interesting to
note that the values of radii for intermediate masses $1.4M_\odot<M<1.6M_\odot$
very weakly depend upon $\alpha$.
Finally, in Sec. \ref{Sec:V} we shall show how for $\alpha<0$ cases, although outside the
star the scalar curvature possess damped oscillations, the gravitational mass contribution when taking into account the
increases with the radial distance.
We conclude the paper by showing our main conclusions in Sec. \ref{Sec:VI}. Figures for all the studied cases are presented at the end of the paper and are intended to help the reader to understand the explained trends in the bulk of the paper.

\section{Equations for compact stars in $f(R)$ gravity}
\label{Sec:II}
Let us consider the gravitational action for $f(R)$ gravity where
$R$ denotes the scalar curvature
\begin{equation}\label{1}
{S}=\frac{1}{16\pi}\int {\rm d}^4 x \sqrt{-g}\,f(R),
\end{equation}
where $G=c=1$ units will be used throughout the paper. The corresponding field equations can be obtained by varying action \eqref{1} with
respect to the metric $g_{\mu\nu}$, yielding
\begin{eqnarray}\label{2}
&&f_{R}G_{\mu\nu}+(\nabla_{\mu}\nabla_{\nu}-g_{\mu\nu}\,\nabla^{\alpha}\nabla_{\alpha})f_{R}-\frac{1}{2}(f(R)-R\,f_{R})\,g_{\mu\nu}\nonumber\\
&&=8\pi T_{\mu\nu}\,,
\end{eqnarray}
where $f_{R}\equiv{\rm d}f(R)/{\rm d}R$,
$G_{\mu\nu}=R_{\mu\nu}-\frac{1}{2}\,R\,g_{\mu\nu}$ is Einstein
tensor and $T_{\mu\nu}$ is energy-momentum tensor, which for a perfect fluid in a comoving frame can be
written  as
\begin{eqnarray}
T_{\mu\nu}=(\rho+p)u_\nu u_\mu-pg_{\mu\nu}\,.
\end{eqnarray}

As easily deduced taking the trace of Eq. \eqref{2}, in GR  ($f(R)\equiv R$)  the scalar curvature is univocally defined by
the energy-momentum content through an algebraic equation. On the contrary, in the context of $f(R)$ theories of gravity in the metric
formalism, given the higher order of the field equations, $R$ may be thought of as an independent dynamical function. 

For a spherically symmetric metric of the form
\begin{equation}\label{40}
{\rm d}s^{2}=-B(r)\,{\rm d}t^{2}+A(r)\,{\rm d}r^{2}+r^{2}({\rm
d}\theta^{2}+\sin^{2}{\theta}\,{\rm d}\phi^{2})\,,
\end{equation}
the corresponding field equations as obtained from Eq. \eqref{2} for functions $A$ and $\psi=B'/B$ (prime denotes derivative with respect to $r$) become
\begin{eqnarray}
\label{5}
A'&=&\frac{2rA}{3f_{R}}\left[8\pi A(\rho+3p)+Af(R)-\,f_{R}\left(\frac{A}{2}R+\frac{3\psi}{2r}\right)\right.\nonumber\\
&&\left.-\left(\frac{3}{r}+\frac{3\psi}{2}\right)f_{2R}\,R'\right]\,,\\
%
\label{6}
\psi'&=&\frac{\psi}{2}\left(\frac{A'}{A}-\psi\right)+\frac{2A'}{rA}+\frac{2}{f_{R}}\left[-8\pi
Ap +\left(\frac{\psi}{2}+\frac{2}{r}\right)\times \right.\nonumber\\
&&\left.f_{2R}\,R' -\frac{A}{2}f(R)\right],\\
%
\label{7}
p'&=&-\frac{\rho+p}{2}\psi\,,
\end{eqnarray}
and the trace of Eq. \eqref{2} gives the equation for the scalar
curvature:
\begin{eqnarray}
\label{8}
R''&=&R'\left(\frac{A'}{2A}-\frac{\psi}{2}-\frac{2}{r}\right)-\frac{A}{3f_{2R}}\left[8\pi(\rho-3p)+f_{R}R\right.\nonumber\\
&&\left. -2f(R)\right]-\frac{f_{3R}}{f_{2R}}\,R'^{2},
\end{eqnarray}
where $f_{2R}={\rm d}^2f(R)/{\rm d}R^2$, $f_{3R}={\rm d}^3f(R)/{\rm d}R^3$.
Finally one needs to specify the $f(R)$ model under consideration
(in our case $f(R)=R+\,\alpha R^2$, with $\alpha$ is constant
parameter) and add the EoS for completeness of the system, i.e., $
p=f(\rho)$. Several EoS will be considered and explained in Sec.
\ref{Sec:III}.
\\

{\it Initial and boundary conditions}: The following conditions at the
centre of star ($r=0$) should be imposed in order to guarantee regularity
and finiteness both of density and pressure everywhere. This provides
\begin{equation}
A(0)=1, \, \psi(0)=0,\ R'(0)=0.
\end{equation}
Therefore there is only one free initial condition for the system of equations Eqs. \eqref{5}-\eqref{8}, namely
the value of the scalar curvature $R(0)$. At the surface of star, both pressure and
density (for most EoS) cancel, i.e., $\rho_{s}=p_{s}=0$. Thus,
from the star radius onwards, one needs to integrate our system of
Eqs. \eqref{5}-\eqref{8} with $\rho=p=0$. The boundary conditions
at spatial infinity can be determined from the
reasonable requirement that the star exterior solution should be matched to
the Schwarzschild one at infinity. Therefore
\begin{equation}
R\rightarrow 0,\quad A\rightarrow 1,\quad r\rightarrow\infty.
\end{equation}
From the equations above, one can see that solutions for $A(r)$,
$\rho(r)$, $p(r)$ and $\psi(r)$ do not depend on the chosen value
$B(0)$. However, in order to achieve, as desirable, Schwarzschild solution at the asymptotic spatial infinity, one must
require $B(r)\rightarrow 1$ for $r\rightarrow \infty$. If
$B(r)\rightarrow B_{\infty}\neq 1$ at $r\rightarrow\infty$, one can
simply rescale the initial condition for $B(0)$ as
\begin{eqnarray}
B(0)\rightarrow \frac{B(0)}{B_{\infty}}.
\end{eqnarray}

On the other hand, on usually assumes $A(r)$ to be of the form
\begin{eqnarray}
A(r)=\left(1-\frac{2m(r)}{r}\right)^{-1}
\end{eqnarray}
and therefore the mass function $m(r)$ as a function of radial coordinate can be expressed as :
\begin{equation}
m(r)=\frac{r}{2}\left(1-A(r)^{-1}\right).
\end{equation}
{\it Mass extraction for negative $\alpha$}:
For the performed analysis with $\alpha<0$, we have seen that $A(r)$ oscillates around unity 
at sufficiently large distances and therefore one cannot extract
the gravitational mass from this parametrization correctly. For
$r\rightarrow \infty$ we have $B\sim {\rm e}^{2\phi}$ where $\phi$
can be interpreted as the Newtonian potential for the
gravitational field at distance $r$, in other words,
$\phi(r)=-m(r)/r$.
Therefore $B(r)\approx 1-\frac{2m(r)}{r}$ at sufficiently large distance $r$. Provided the gravitational
mass has a well-defined limit one can assume that $m(r)\approx
M=const$ from some $r=r_{c}$ onwards ($r_{c}$ being a sufficiently
large distance) and thus, the gravitational mass $M$ can be extracted from $B(r)$. Consequently for an initial
value of $B(0)$ so that $B(r)\rightarrow 1$ at
$r\rightarrow\infty$ we have
\begin{eqnarray}
B(r)=1-\frac{2m(r)}{r}\,,
\end{eqnarray}
and therefore the gravitational mass $M$ can be calculated as the
limit
\begin{equation}\label{mass1}
M\equiv \lim_{r\rightarrow\infty}r(1-B(r))/2.
\end{equation}
{\it Mass extraction for positive $\alpha$}:
As we shall see in Sec. \ref{Sec:IV}, in the event $\alpha>0$, the limit $m(r)\rightarrow M$ at spatial infinity is well
defined. In the event that the value of $B(0)$ is arbitrarily chosen, one first needs to obtain the asymptotic value
$B_{\infty}=\lim_{r\rightarrow\infty}B(r)$, so that function $B(r)$ becomes
\begin{eqnarray}
B(r)=B_{\infty}\left(1-\frac{2m(r)}{r}\right)\,,
\end{eqnarray}
so a simple relation for gravitational mass
\begin{eqnarray}\label{mass2}
M=\frac{1}{B_{\infty}}\lim_{r\rightarrow\infty}\frac{r(B_{\infty}-B(r))}{2}\,,
\end{eqnarray}
is obtained.

We need carefully investigate the possibility of appearance of
contributions with $r^{\gamma}$ ($\gamma>0$) in $B(r)$. In this
case we have no finite limit for mass.
The behaviour of the resulting solution for the metric functions $A$ and $B$ in the
outer region depends values $A(r)$, $B(r)$ and
its derivatives on star surface ($r=r_s$). For the latter, specific values depend upon the chosen EoS.

\section{Realistic equations of state}
\label{Sec:III}
For our investigation we have considered a representative set
of four realistic EoS.

{\it Neutron stars}: Below we provide the main features and assumptions in the EoS under study herein.
\begin{itemize}

\item[$1)$] the well-known SLy4 EoS \cite{SLy, SLy-4} is obtained from
many body calculations with simple two-nucleon potential. For completeness, we have
also considered the AP4 EoS \cite{APR} which hosts a
three-nucleon potential and Argonne 18 potential with UIX
potential;

\item[$2)$] 
as an example of EoS based upon the relativistic mean-field (RMF) calculations we take the generalisation of a model firstly considered by Glendenning and Moszkowski (GM) 
\cite{GM}. This generalisation, dubbed GM1nph, has been proposed for cold neutron star matter in $\beta$-equilibrium containing the  baryon octet and electrons \cite{GM-1};

\item[$3)$] relatively stiff MPA1 EoS \cite{MPA} obtained from relativistic
Dirac-Brueckner-Hartree-Fock formalism. This EoS accounts for the energetic
contributions originated by the exchange between $\pi$ and $\rho$ mesons.
\end{itemize}

\begin{figure}[H]\begin{center}
  \includegraphics[scale=0.85]{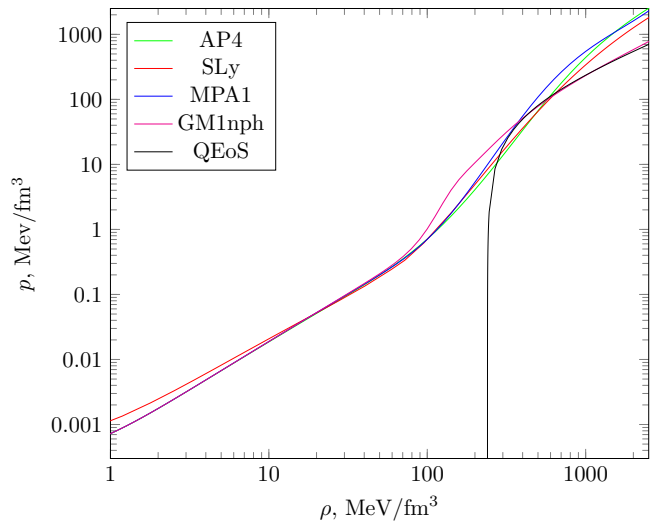}
\caption{Equations of state for nuclear matter (AP4, SLy, MPA1,
GM1nph) and for quark matter (QEoS).} \label{Fig1}
\end{center}
\end{figure}

On Fig. \ref{Fig1} we plot dependence of pressure from energy
density for these four EoS. One can see that for
intermediate densities these EoS seem indistinguishable. For GM1
EoS there is a softening effect at high densities due to appearance of
hyperons (mainly $\Sigma^{-}$, $\Lambda^{0}$ and $\Sigma^{0}$).
In our calculations we have used analytical
representations of these EoS. Thus, the SLy EoS can be represented
as
\begin{eqnarray}
\label{FPS}
\zeta&\,=\,&\frac{a_{1}+a_{2}\xi+a_{3}\xi^3}{1+a_{4}\xi}K(a_{5}(\xi-a_{6}))+(a_{7}+a_{8}\xi)\times\nonumber\\
&&K(a_{9}(a_{10}-\xi))+(a_{11}+a_{12}\xi)K(a_{13}(a_{14}-\xi))\nonumber\\
&&+(a_{15}+a_{16}\xi)K(a_{17}(a_{18}-\xi)),
\end{eqnarray}
where $\zeta\equiv\log(P/{\rm dyn}\,{\rm cm}^{-2})$,
$\xi\equiv\log(\rho/\mbox{g}\,\mbox{cm}^{-3})$, and  $K(x)\equiv
1/({\rm e}^x+1)$.
The coefficients $a_{i}$ can be found in \cite{Camenzind}. For
MPA1, AP4 and GM1nph EoS take the more complicated dependence
shown in \cite{Eksi}, namely
\begin{eqnarray}
\label{another}
\zeta=\zeta_{low}K(a_{1}(\xi-c_{11}))+\zeta_{high}K(a_{2}(c_{12}-\xi)),
\end{eqnarray}
with
\begin{eqnarray}
\zeta_{low}&=&\left[c_1+c_2(\xi-c_3)^{c_4}\right]K(c_{5}(\xi-c_6))\nonumber\\
&&+\,(c_7+c_6\xi)K(c_{9}(c_{10}-\xi)),\\
\zeta_{high}&=&(a_{3}+a_{4}\xi)K(a_{5}(a_{6}-\xi))\nonumber\\
&&+\,(a_7+a_8\xi+a_{9}\xi^2)K(a_{10}(a_{11}-\xi))\,,
\end{eqnarray}
where $a_{i}$ and $c_{i}$ are constant coefficients provided in \cite{Eksi}.

{\it Quark stars}: Apart from the EoS described above, we have also considered the case of 
quark (or so called strange) stars \cite{Itoh, Witten}. It is assumed that
quark stars consist of $u$, $d$ and $s$ quarks and electrons. The
deconfined quarks can form a colour superconductor. This leads to
a softer EoS with possible observable effects on the
minimum allowed mass, radii, cooling behaviour and other observables.
One of the simplest EoS for quark matter is provided by the so-called {\it MIT
bag} model \cite{Jaffe, Simonov} as follows:
\begin{equation}
\label{QEOS} p=b(\rho-4B)\,,
\end{equation}
where $B$ is the bag constant. The value of parameter $b$ depends
on both the chosen mass for the strange quark $m_s$ and QCD
coupling constant and usually varies from $b=1/3$ ($m_s=0$) to
$b=0.28$ ($m_{s}=250$ MeV). The value of $B$ lies in interval
$0.98<B<1.52$ in units of $B_{0}=60$ MeV/fm$^{3}$ \cite{Sterg}. In
the following, we shall consider the case $B=1$ and $b=0.31$.

\section{Results for $\alpha>0$}
\label{Sec:IV}
\begin{figure*}
\begin{center}
  \includegraphics[scale=0.7]{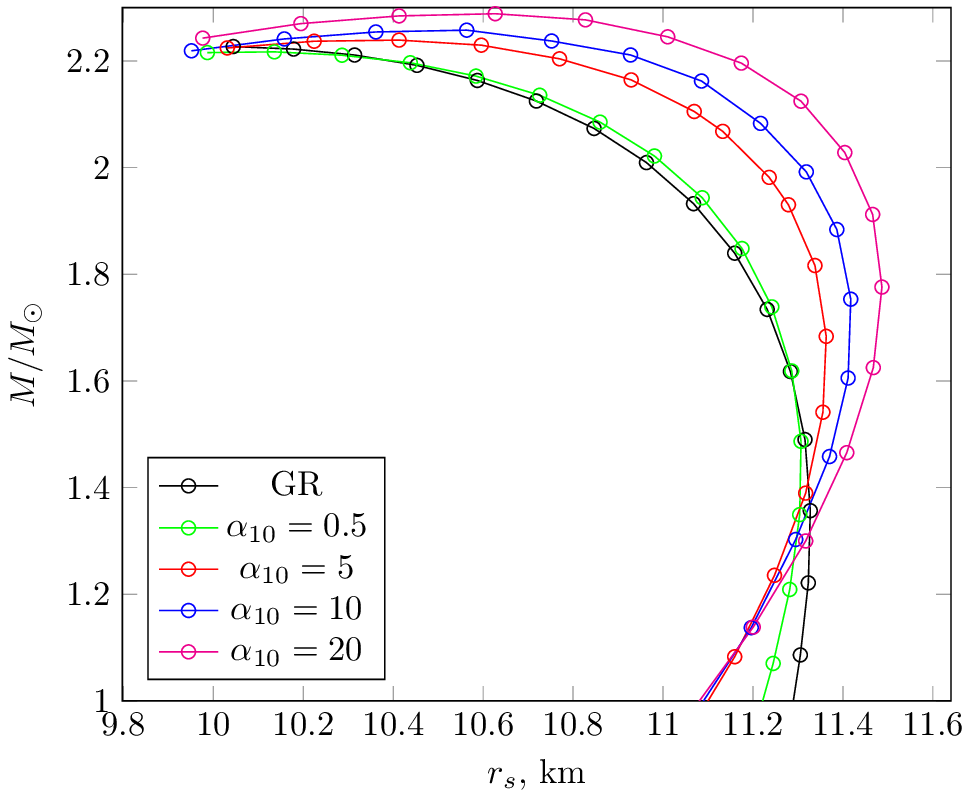}\includegraphics[scale=0.7]{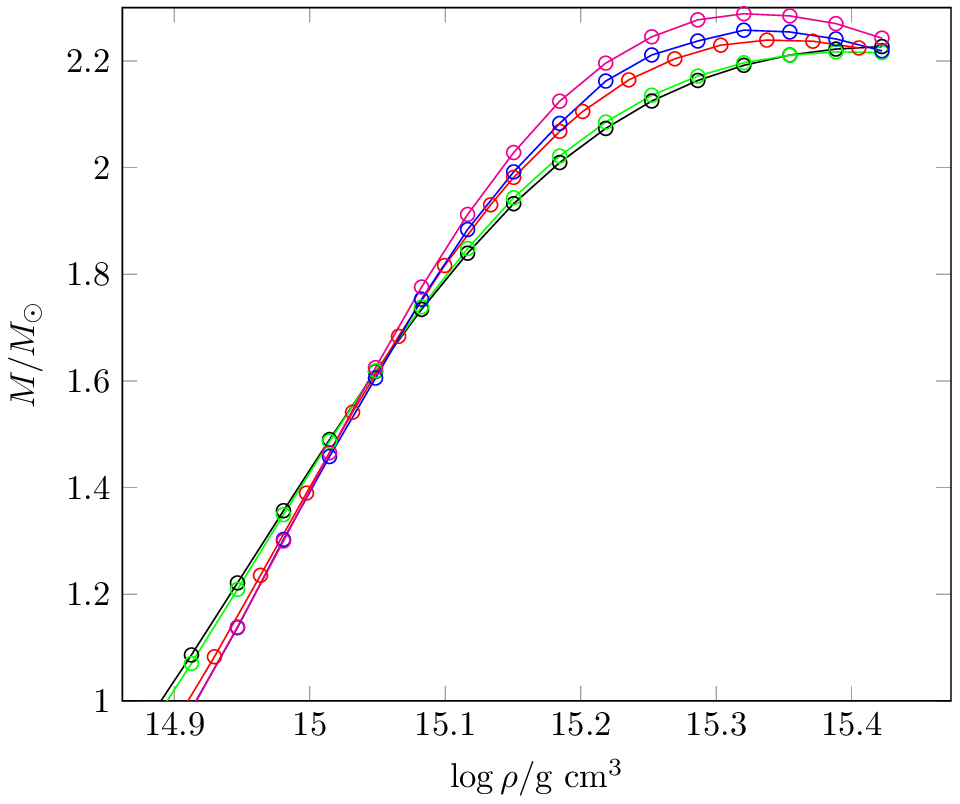}\\
  \includegraphics[scale=0.7]{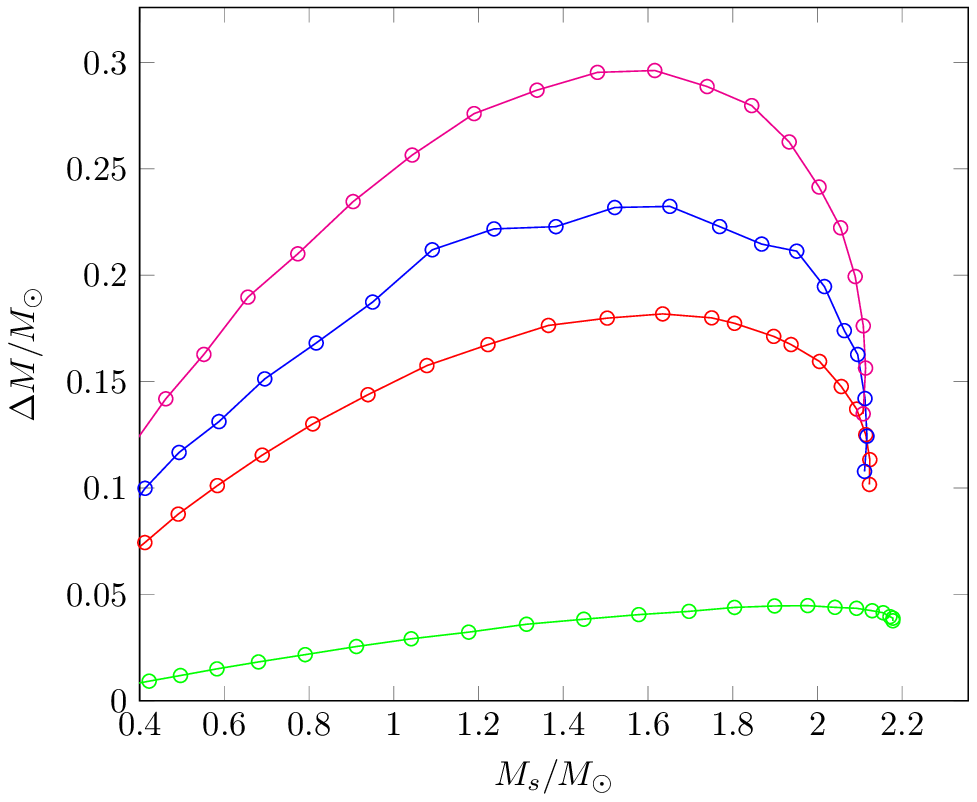}\includegraphics[scale=0.7]{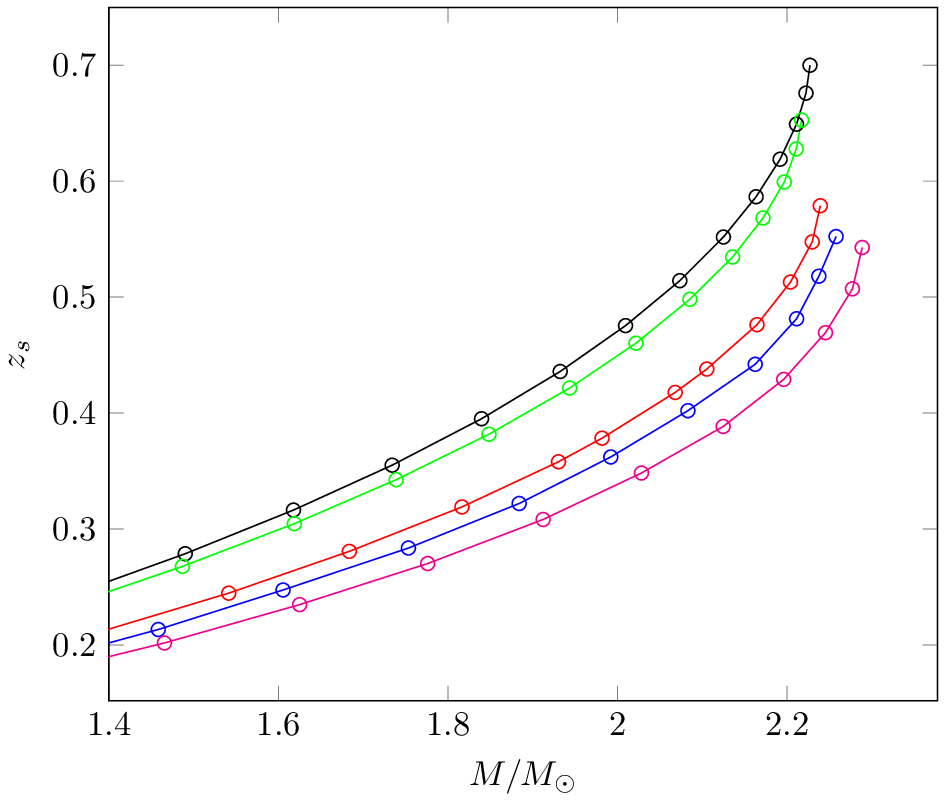}
\caption{Upper panel: relation between the gravitational mass
($M$) and radius of the star for AP4 EoS in $R^2$-gravity when compared  with the
General Relativity predictions (left); relation between the gravitational mass and the central density
of the star (right). The symbol $\alpha_{10}$ refers to $\alpha\times 10^{10}$
cm$^2$. Lower panel: the dependence of the contribution of
gravitational sphere into gravitational mass $\Delta M=M-M_{s}$
with stellar mass $M_{s}\equiv m(r_{s})$ (left); surface redshift as a
function of gravitational mass (right).}
\label{Fig2}
\end{center}
\end{figure*}

Following the procedure described in Sec. \ref{Sec:II} we have calculated the mass-radius and mass-central density 
relations for some values of parameter $\alpha$ for all EoS
explained in Sec.
\ref{Sec:III} above. Results are depicted in Figs. \ref{Fig2}, \ref{Fig4}, \ref{Fig6}, \ref{Fig8} and \ref{Fig10}. 
%
More specifically, for a given initial pressure, we have integrated the system of Eqns. \eqref{5}-\eqref{8} using the EoS described in Sec. \ref{Sec:III}. As soon as $p(r=r_s)=0$, we label that radial coordinate value as the star radius $r_s$. From this value onwards, obviously we are in vacuum and the aforementioned system of equations is integrated with zero density and zero pressure. The determination of the asymptotic mass $M$ follows the procedure described in \ref{Sec:III}, so a pair $\{M_,\,r_s\}$ is found, which gives rise to results in Figs. \ref{Fig2}, \ref{Fig4}, \ref{Fig6}, \ref{Fig8} and \ref{Fig10}. One needs to be aware that as in the GR case, not every initial pressure value is able to host a stable star, so in this sense every pair $\{M_
,\,r_s\}$ corresponds to a different initial pressure. 

As has been explained in the Introduction, in the context of $f(R)$ extended
theories of gravity, one needs to distinguish between stellar mass bounded by
star surface ($M_{s}=m(r_{s})$) and gravitational mass as measured by a distant observer ($M=m(r\rightarrow\infty)$). Obviously, in GR these two values coincide, but in $R^2$-gravity
the {\it gravitational sphere} around the star emerges and contributes
 to the gravitational mass defined as
$M(r_{\infty})$. We have observed that for the same central
density, $R^2$ stellar masses $M_s$ are smaller than their GR
counterparts. This decrease is then partially compensated by
energetic contribution of the {\it gravitational sphere} when the
gravitational mass $M$ is computed. Consequently, the resulting
gravitational mass-radius relation ends up to differ from the GR
prediction. The {\it effective radius} of the {\it gravitational
sphere} can be determined as the radius at which the
metric solution is sufficiently close to the Schwarzschild solution (for
example the radius at which the scalar curvature is of the order of magnitude $R \sim
10^{-5}$ in units of $r_{g}^{-2}\equiv c^4/G^2M^2_\odot$). We have observed that the
{\it effective radius} turns out to be proportional to $\alpha^{1/2}$.
\\

\begin{figure*}
\begin{center}
\includegraphics[scale=0.7]{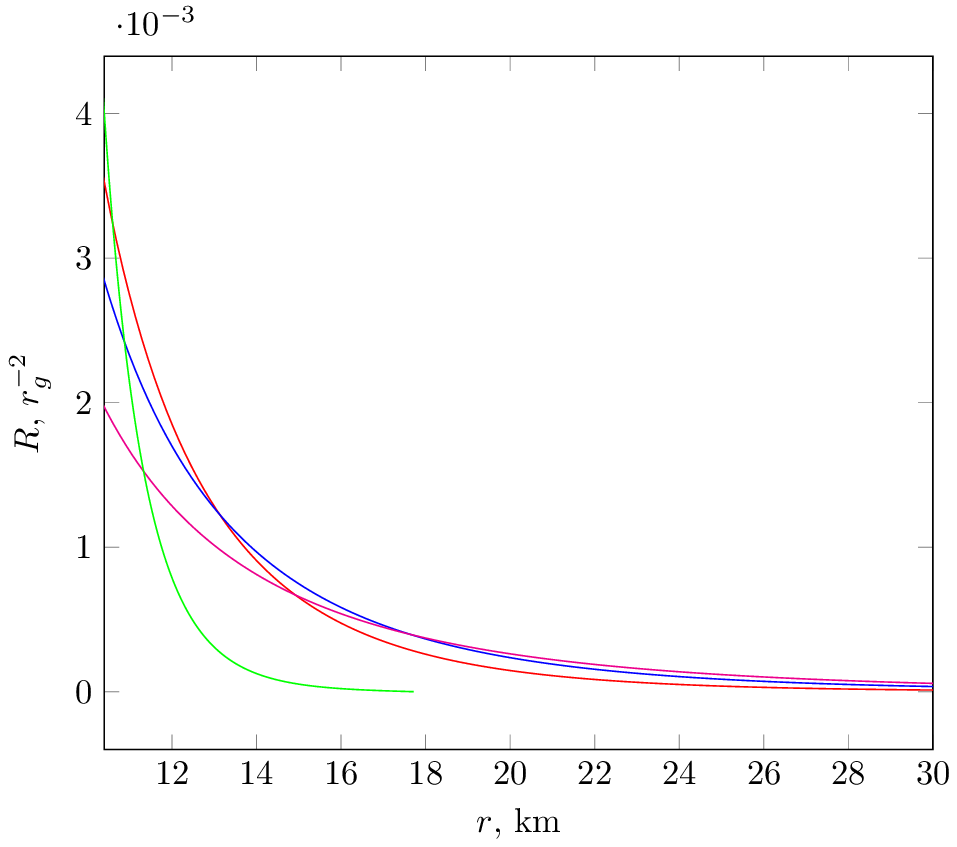}\includegraphics[scale=0.7]{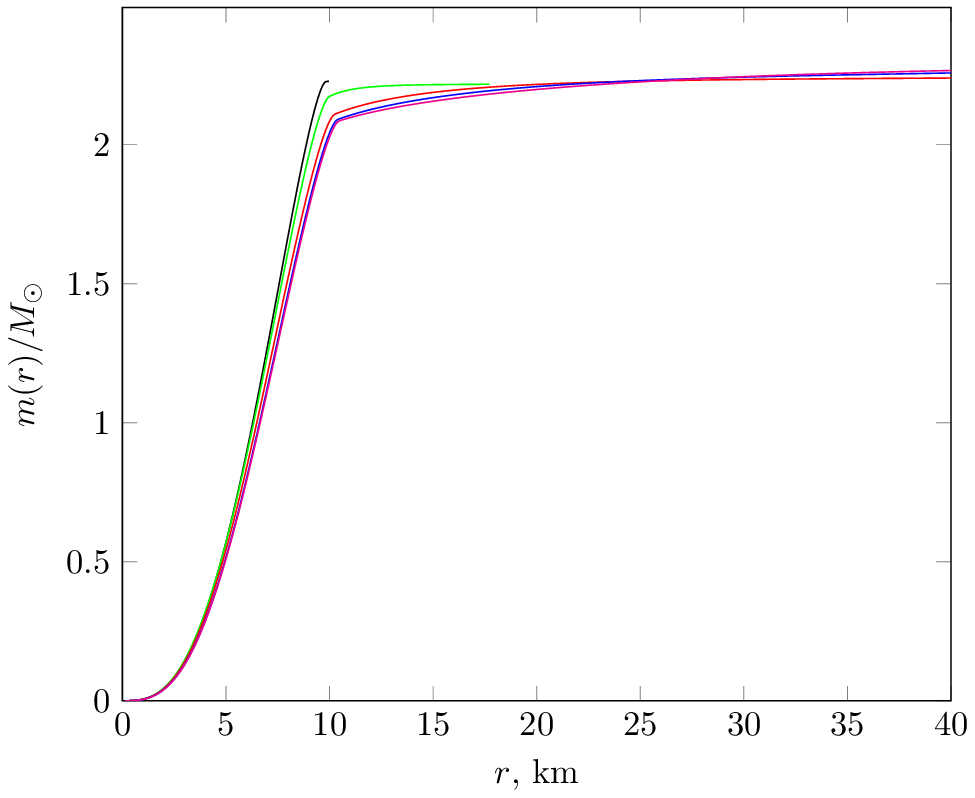}
\caption{Left panel: Dependence of the scalar curvature outside
the star for stellar configurations with maximal mass for AP4 EoS.
Scalar curvature is given in units of $r_{g}^{-2}\equiv
c^4/G^2M^2_\odot$). Right panel: the stellar mass profile $m(r)$
for stellar configurations with maximal gravitational mass.}
\label{Fig3}
\end{center}
\end{figure*}

\begin{figure*}
\begin{center}
  \includegraphics[scale=0.7]{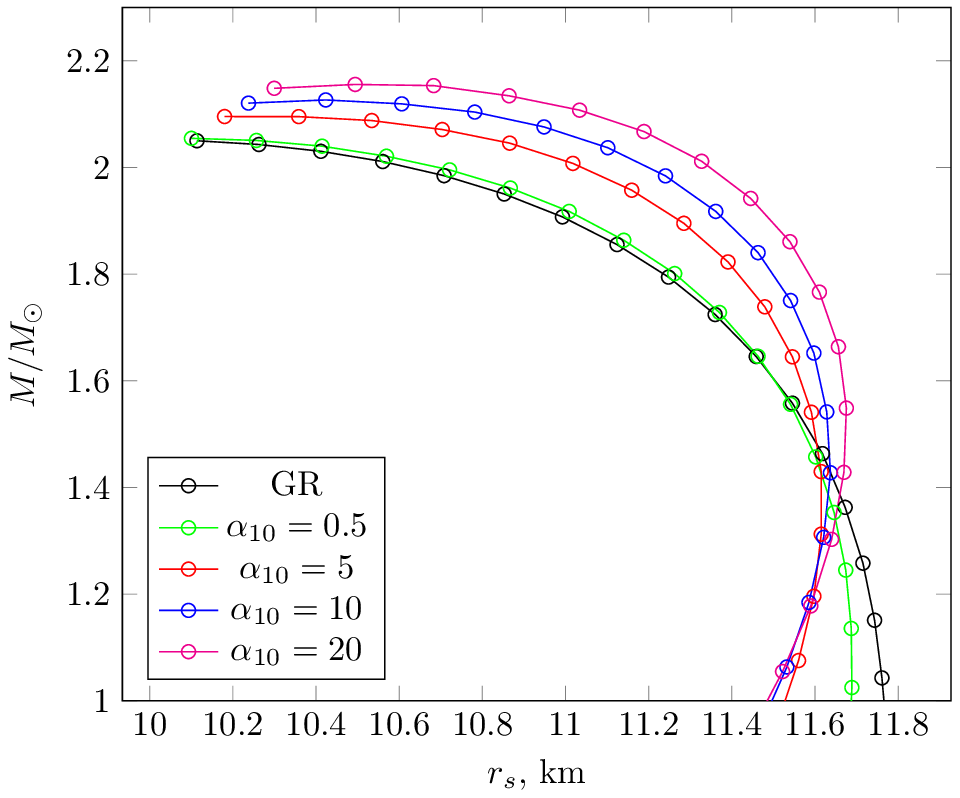}\includegraphics[scale=0.7]{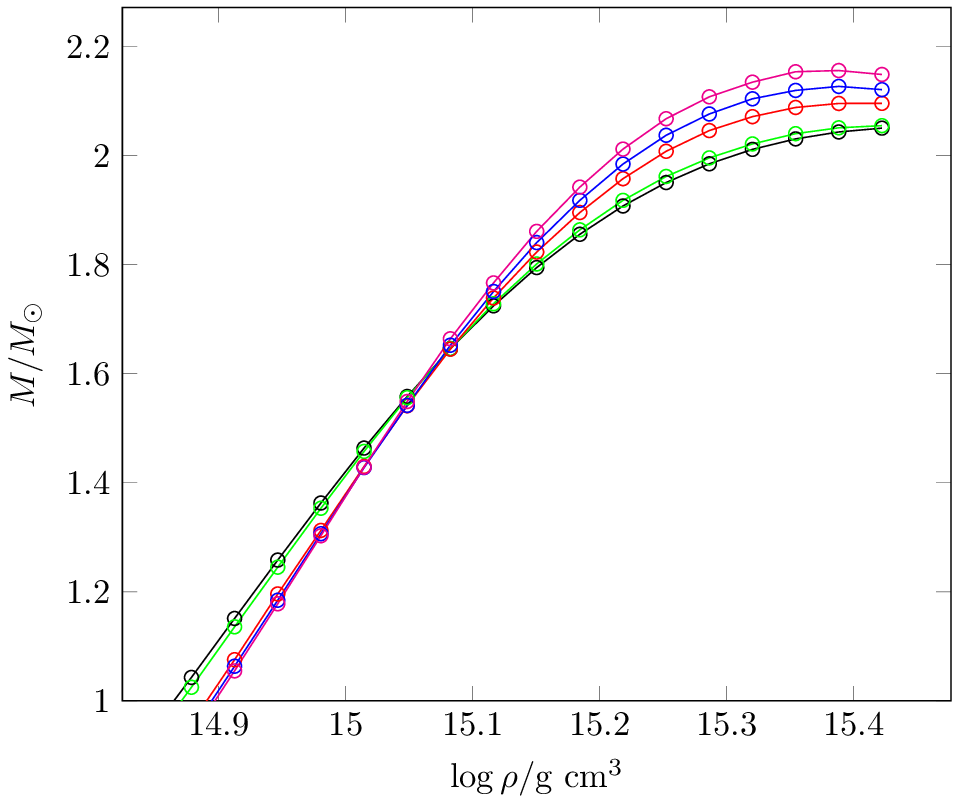}\\
  \includegraphics[scale=0.7]{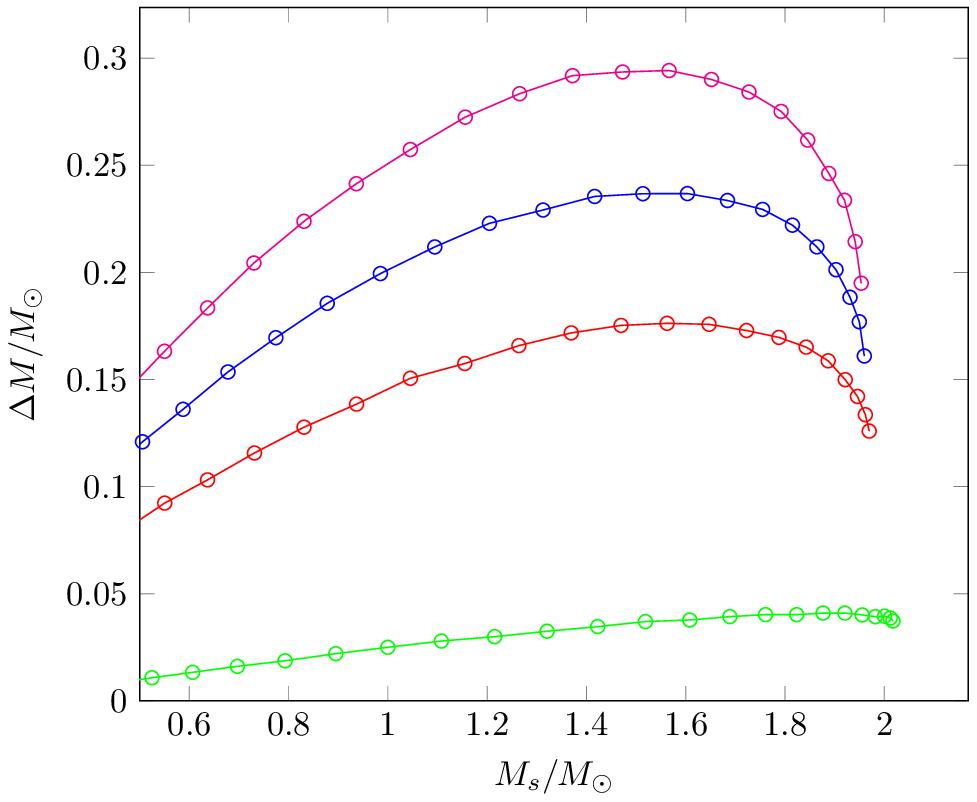} \includegraphics[scale=0.7]{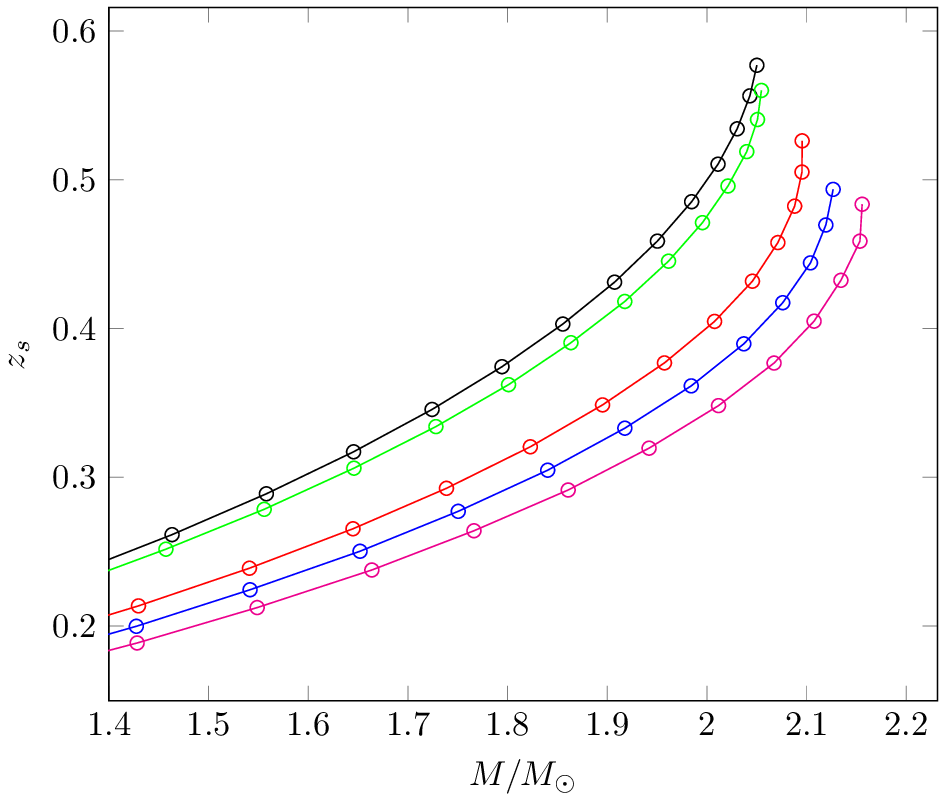}
\caption{The same as on Fig. \ref{Fig2} but for SLy EoS.}
\label{Fig4}
\end{center}
\end{figure*}

\begin{figure*}
\begin{center}
\includegraphics[scale=0.7]{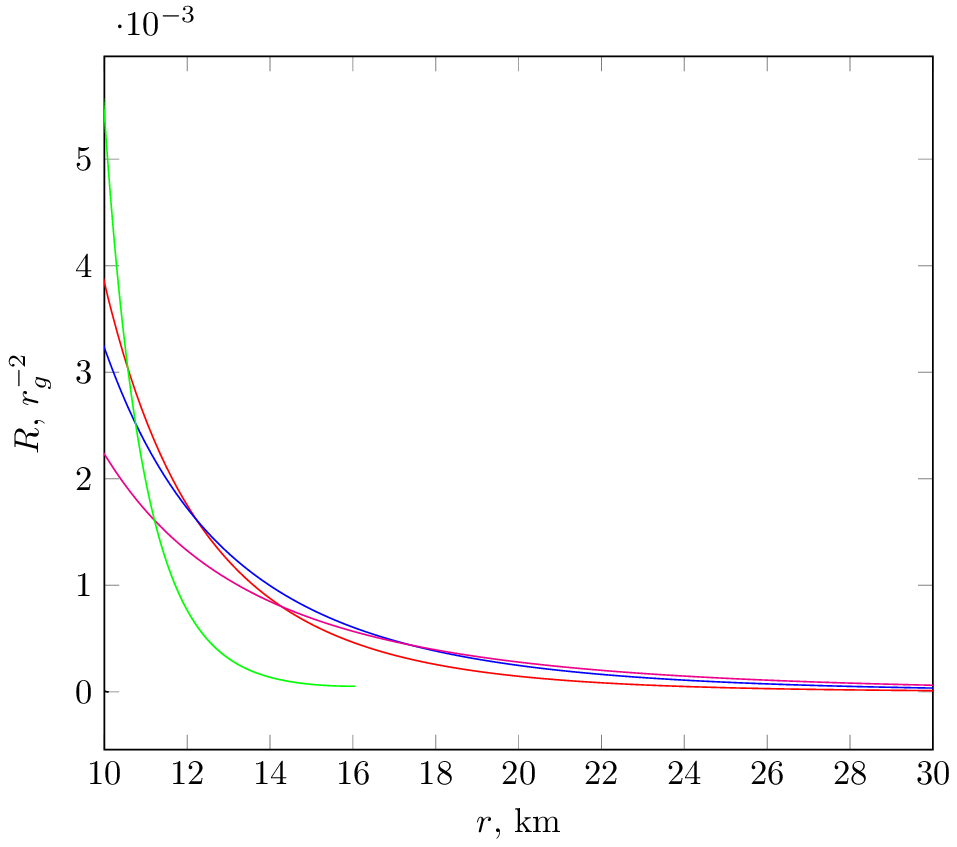}\includegraphics[scale=0.7]{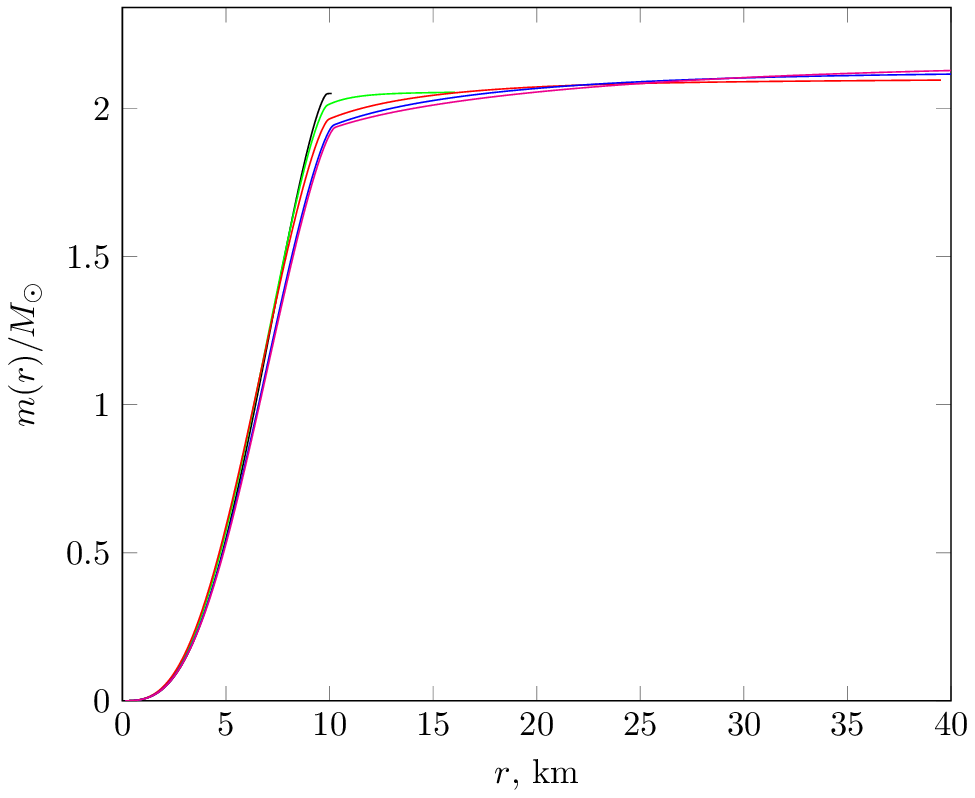}
\caption{The same as on Fig. \ref{Fig3} but for SLy EoS.}
\label{Fig5}
\end{center}
\end{figure*}

\begin{figure*}
\begin{center}
  \includegraphics[scale=0.7]{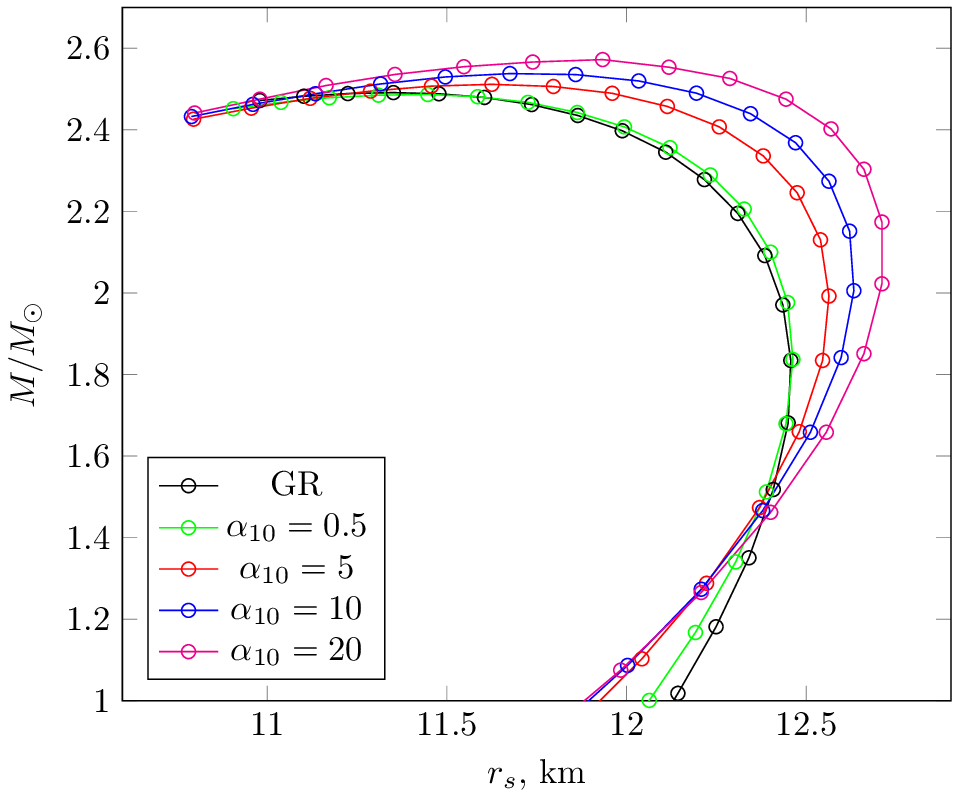}\includegraphics[scale=0.7]{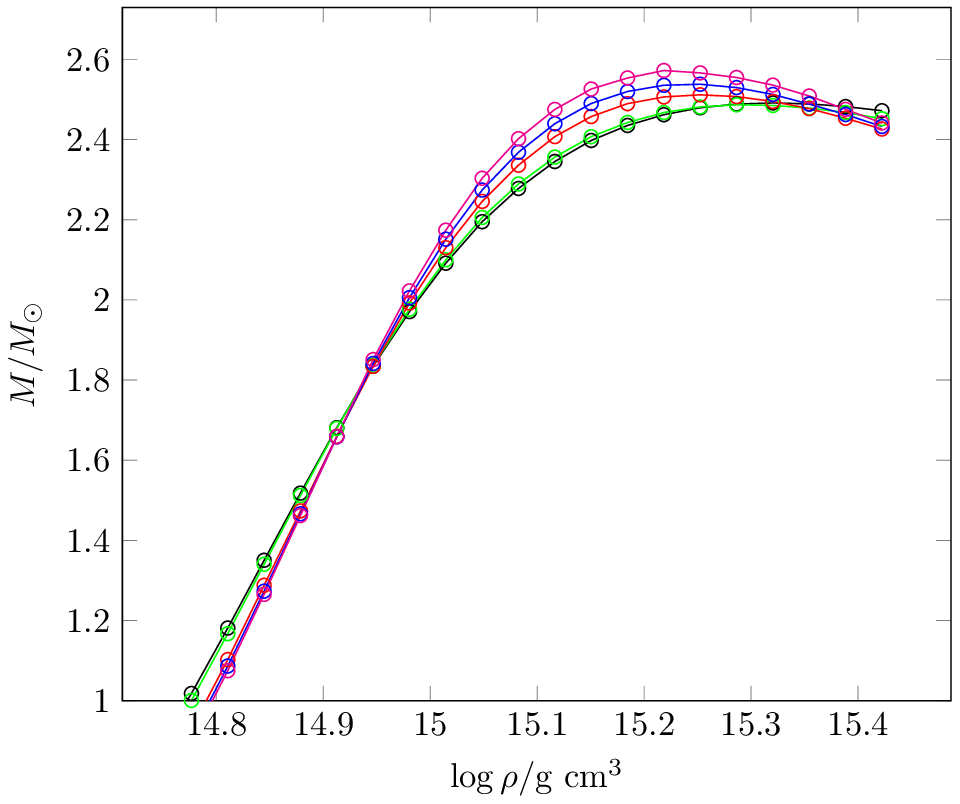}\\
  \includegraphics[scale=0.7]{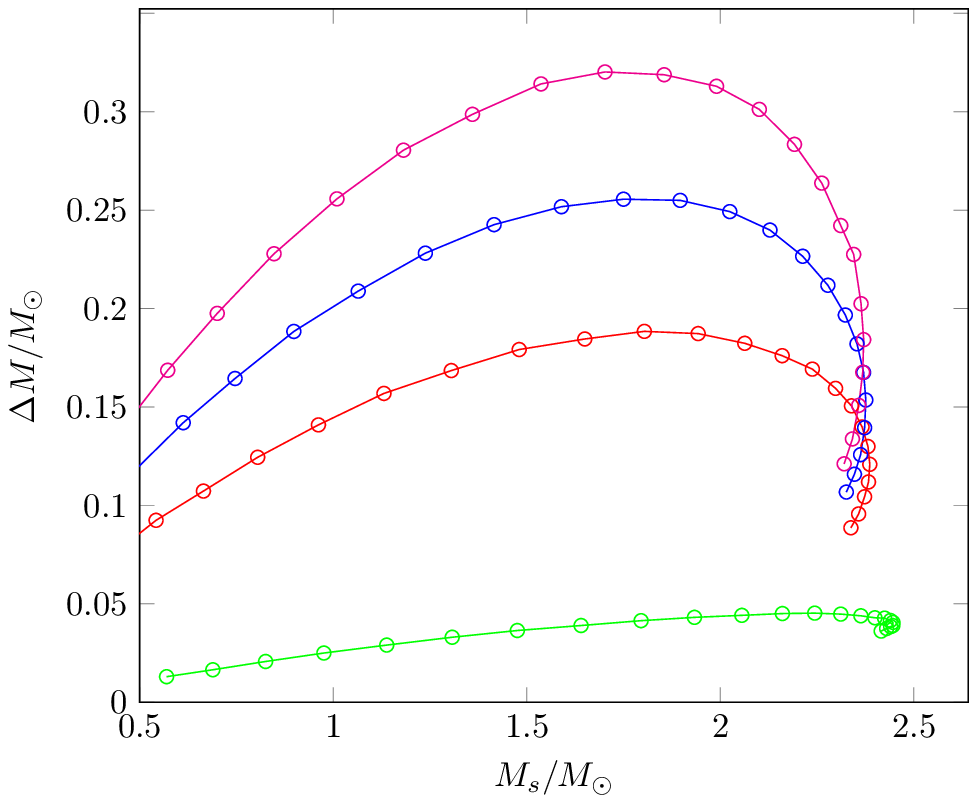}\includegraphics[scale=0.7]{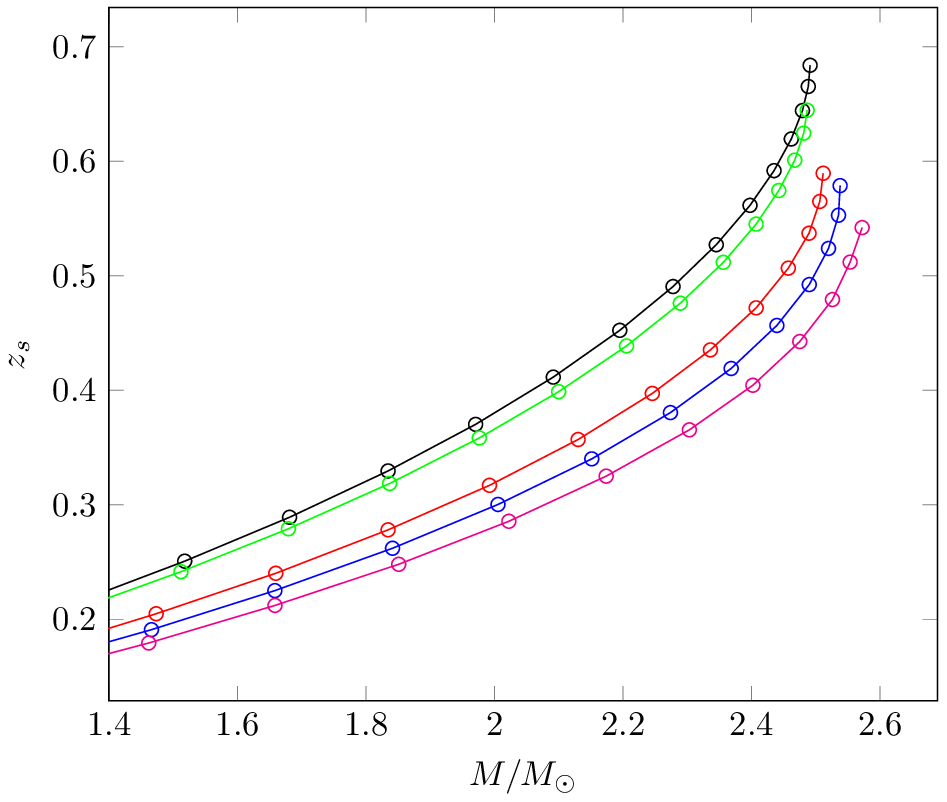}
\caption{The same as on Fig. \ref{Fig2} but for stiff MPA1
EoS.}
\label{Fig6}
\end{center}
\end{figure*}

\begin{figure*}
\begin{center}
\includegraphics[scale=0.7]{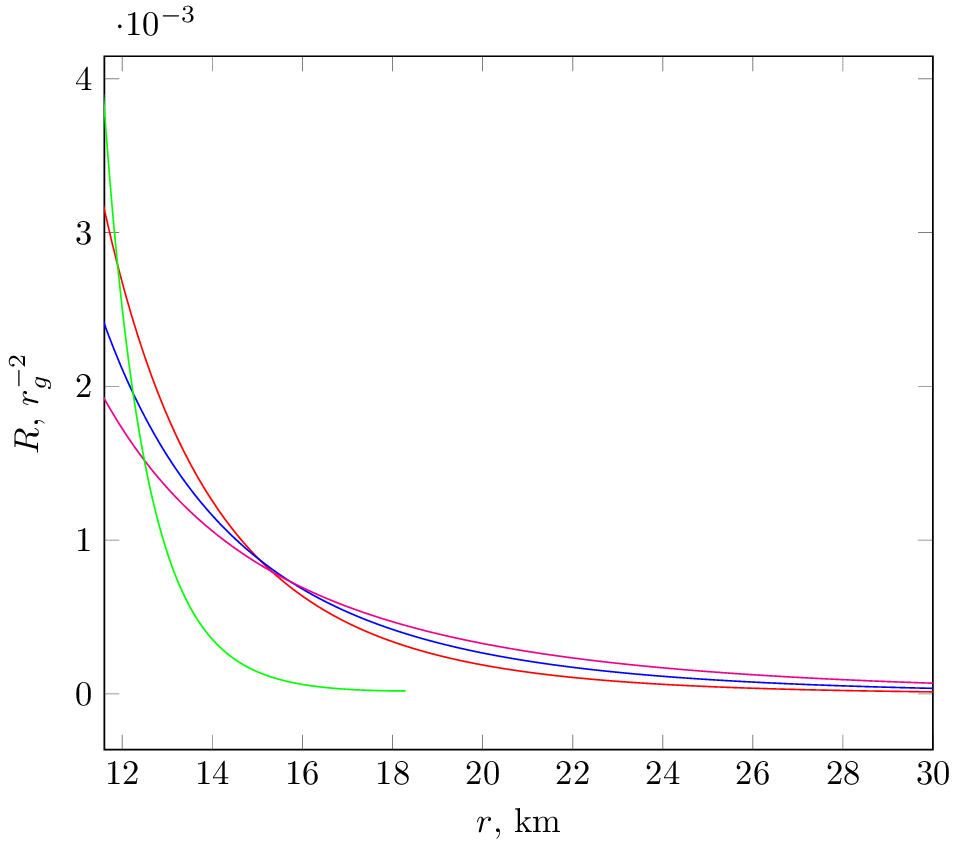}\includegraphics[scale=0.7]{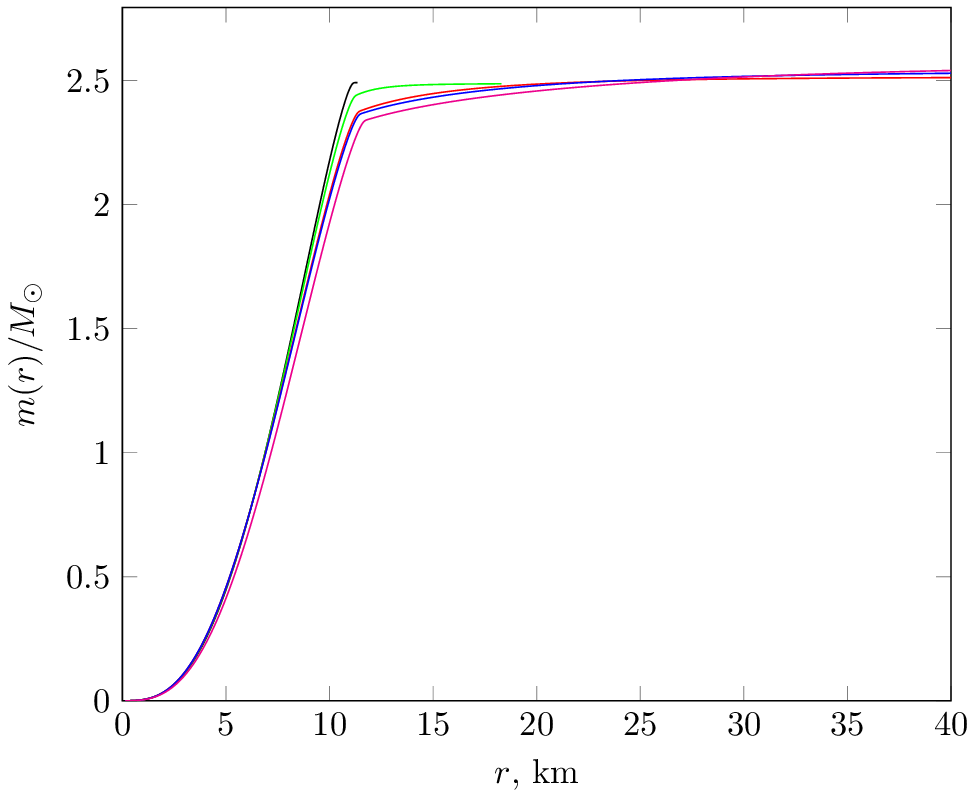}
\caption{The same as on Fig. \ref{Fig3} but for stiff MPA1
EoS.}
\label{Fig7}
\end{center}
\end{figure*}

\begin{figure*}
\begin{center}
  \includegraphics[scale=0.7]{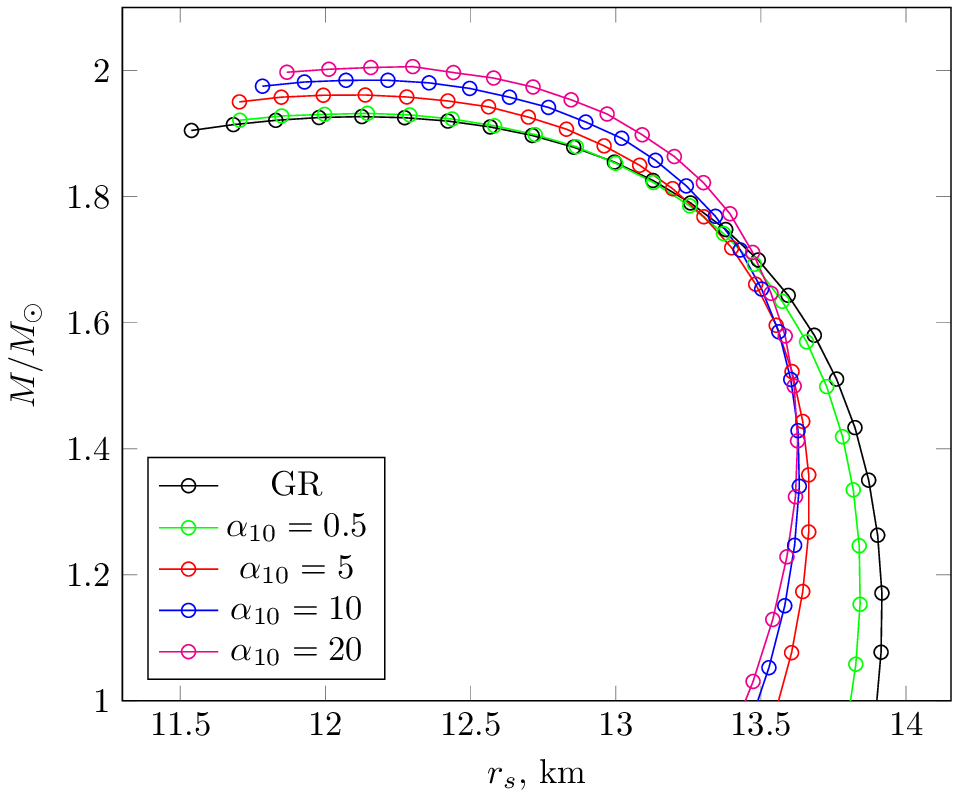}\includegraphics[scale=0.7]{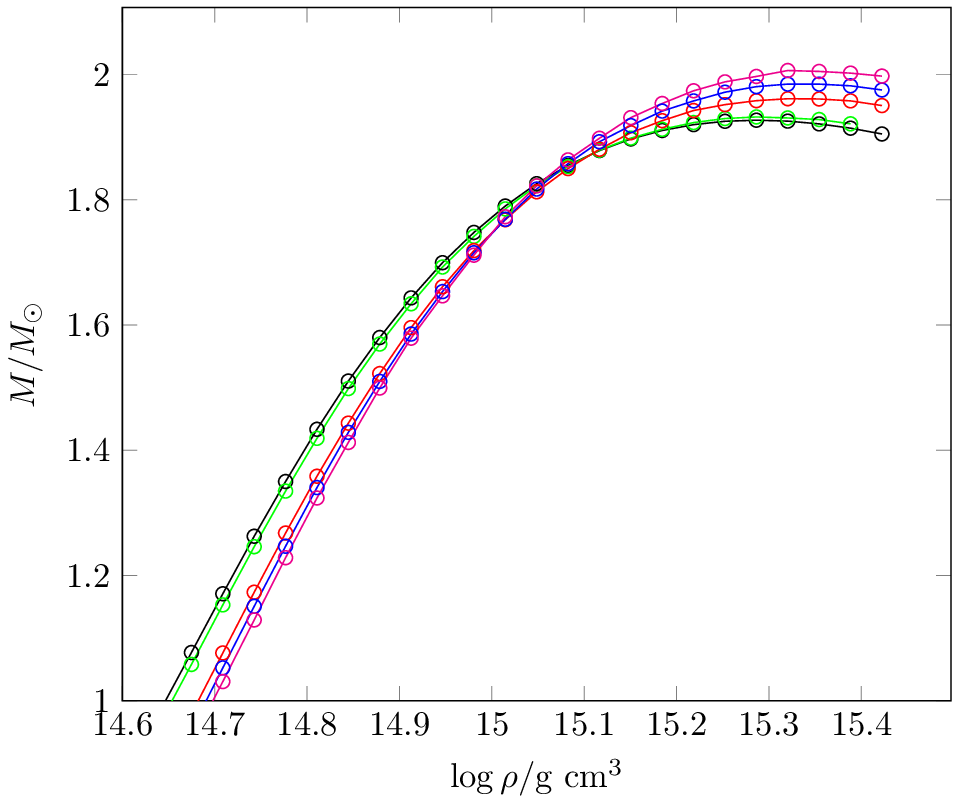}\\
  \includegraphics[scale=0.7]{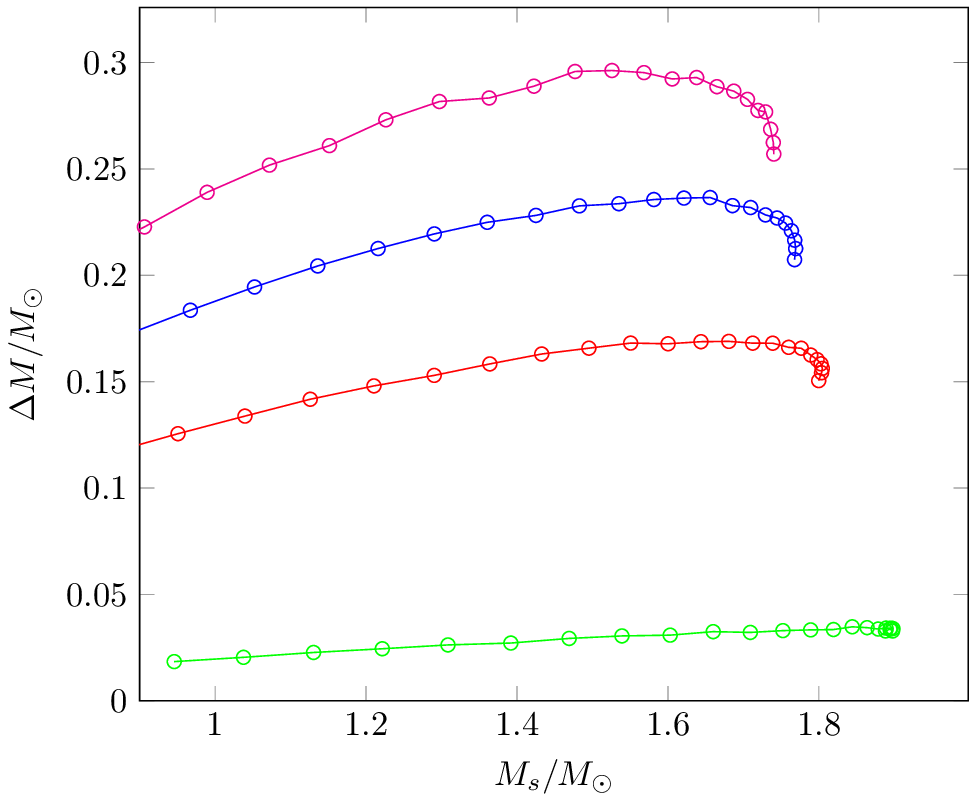}\includegraphics[scale=0.7]{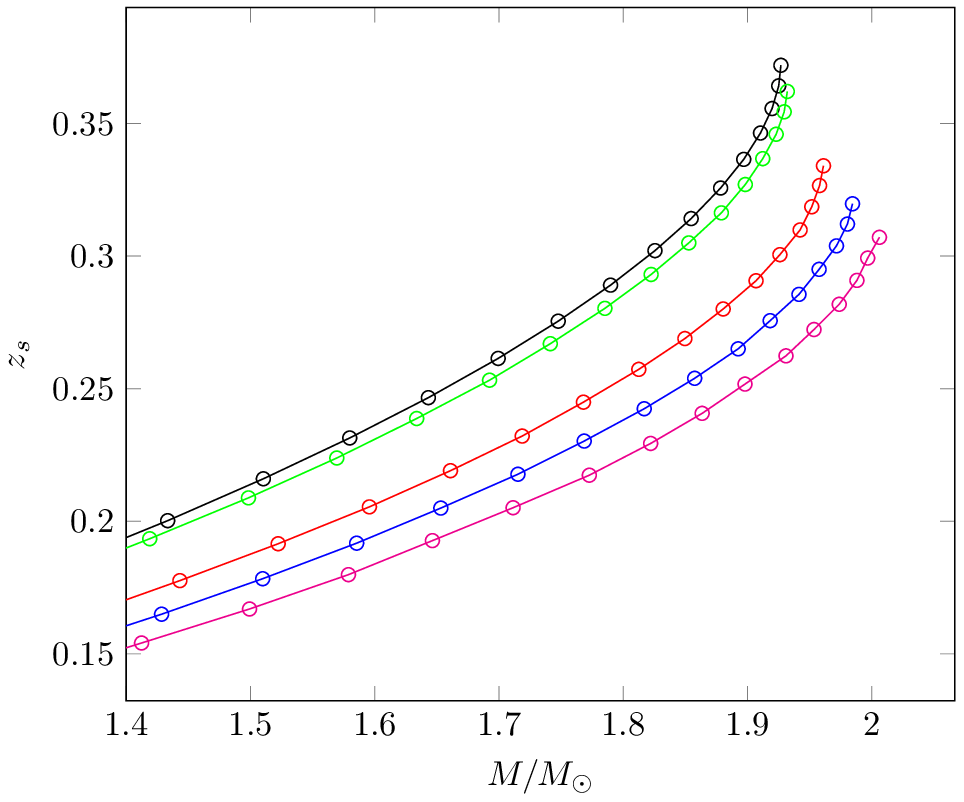}
\caption{The same as on Fig. \ref{Fig2} but for GM1 EoS with inclusion of
hyperons.}
\label{Fig8}
\end{center}
\end{figure*}

\begin{figure*}
\begin{center}
\includegraphics[scale=0.7]{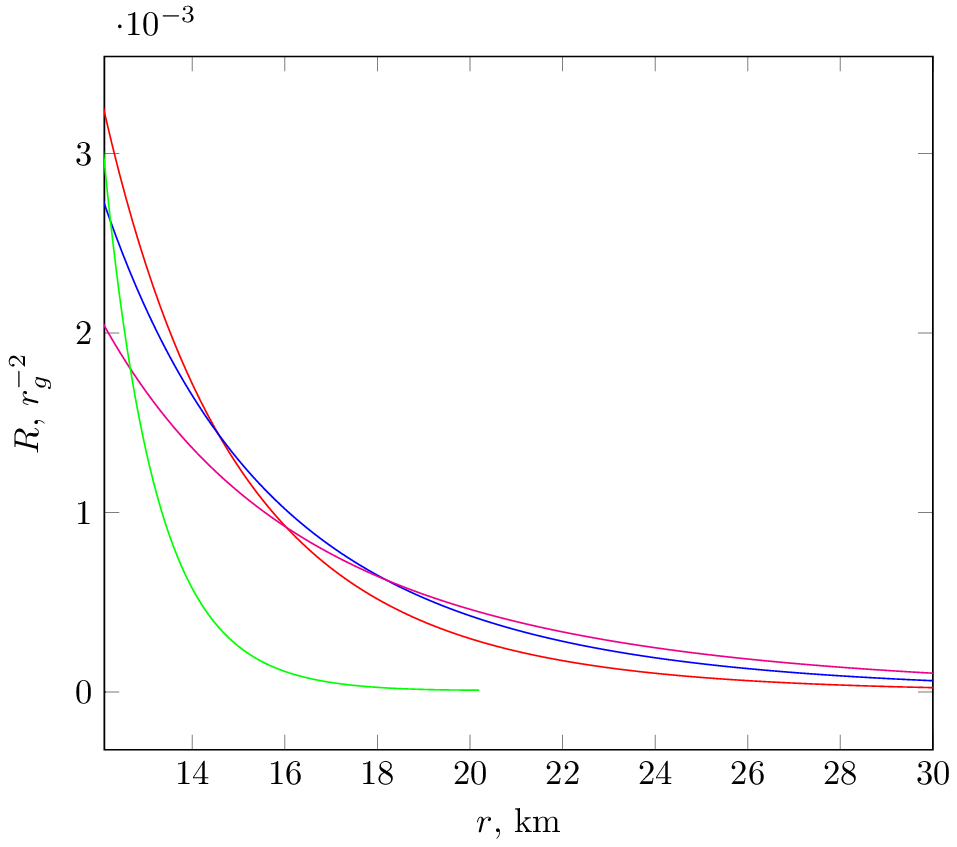}\includegraphics[scale=0.7]{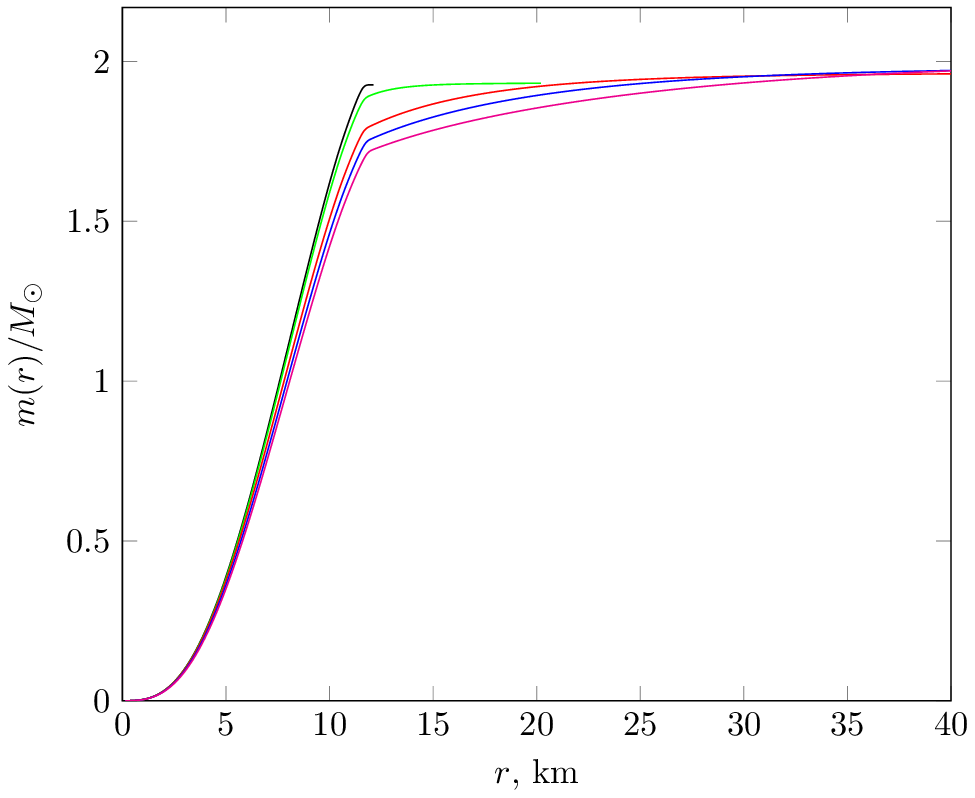}
\caption{The same as on Fig. \ref{Fig3} but for GM1 EoS with inclusion of
hyperons.}
\label{Fig9}
\end{center}
\end{figure*}

\begin{figure*}
\begin{center}
  \includegraphics[scale=0.7]{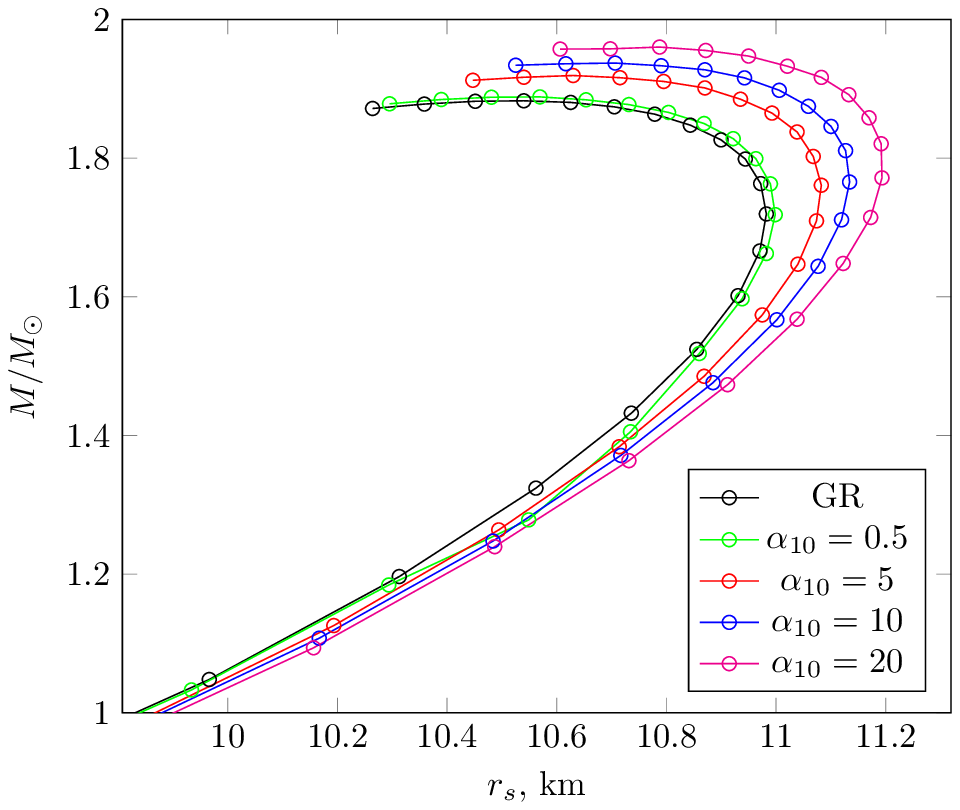}\includegraphics[scale=0.7]{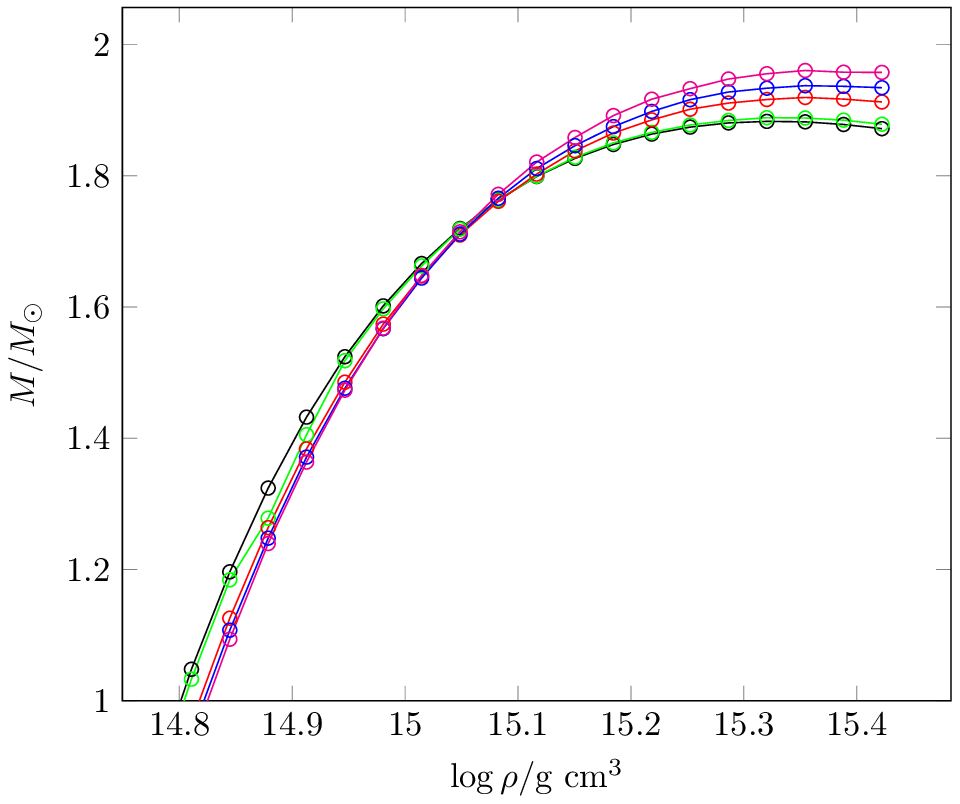}\\
  \includegraphics[scale=0.7]{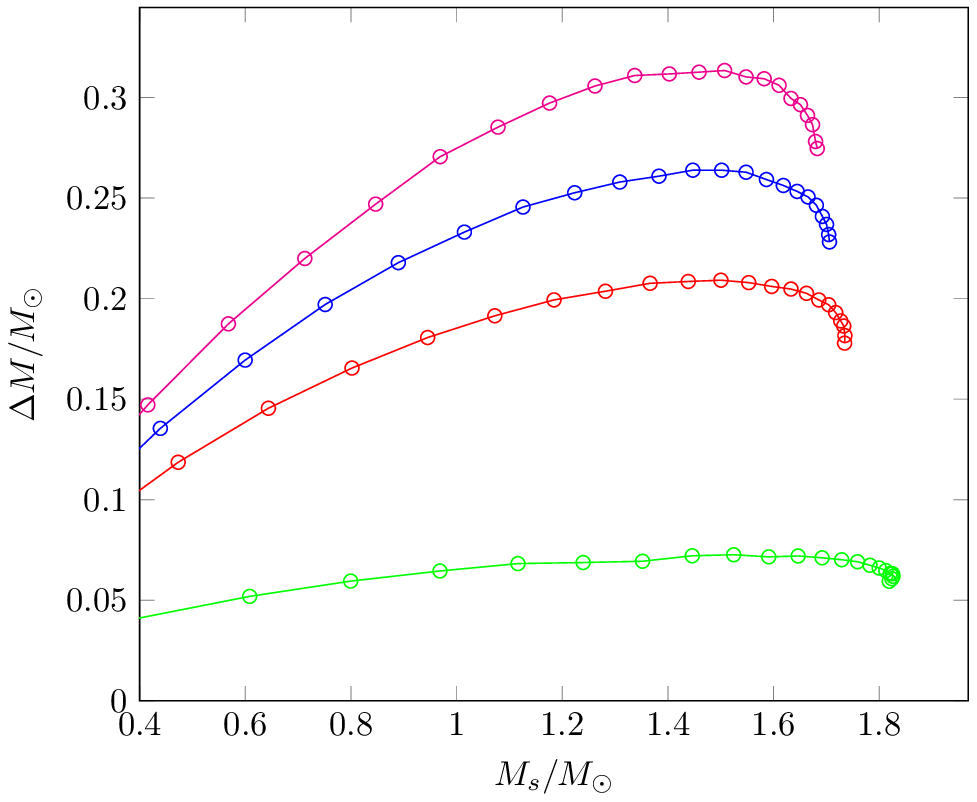}\includegraphics[scale=0.7]{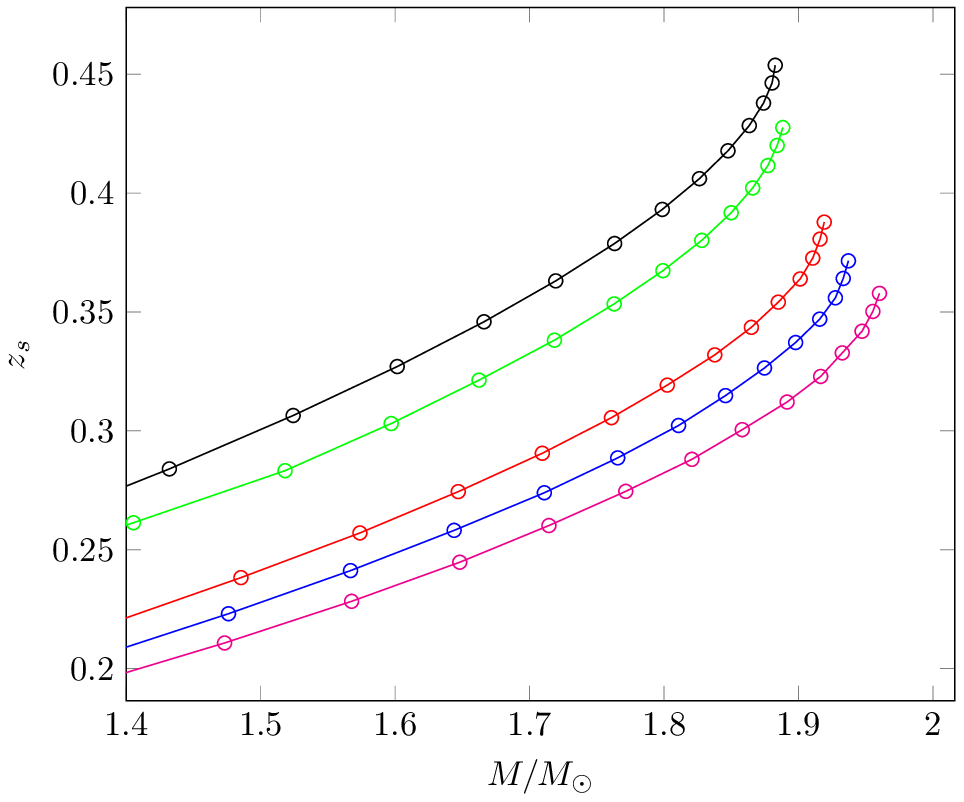}
\caption{Upper panel: relation between gravitational mass $M$
and radius for simple quark EoS for $R^2$-gravity in comparison
with General Relativity (left); relation between gravitational mass and central density
of the star. Lower panel: the dependence of the contribution
of gravitational sphere into gravitational mass $\Delta M=M-M_{s}$
from $M_{s}$ (left); surface redshift as a
function of gravitational mass (right).}
\label{Fig10}
\end{center}
\end{figure*}

\begin{figure*}
\begin{center}
\includegraphics[scale=0.7]{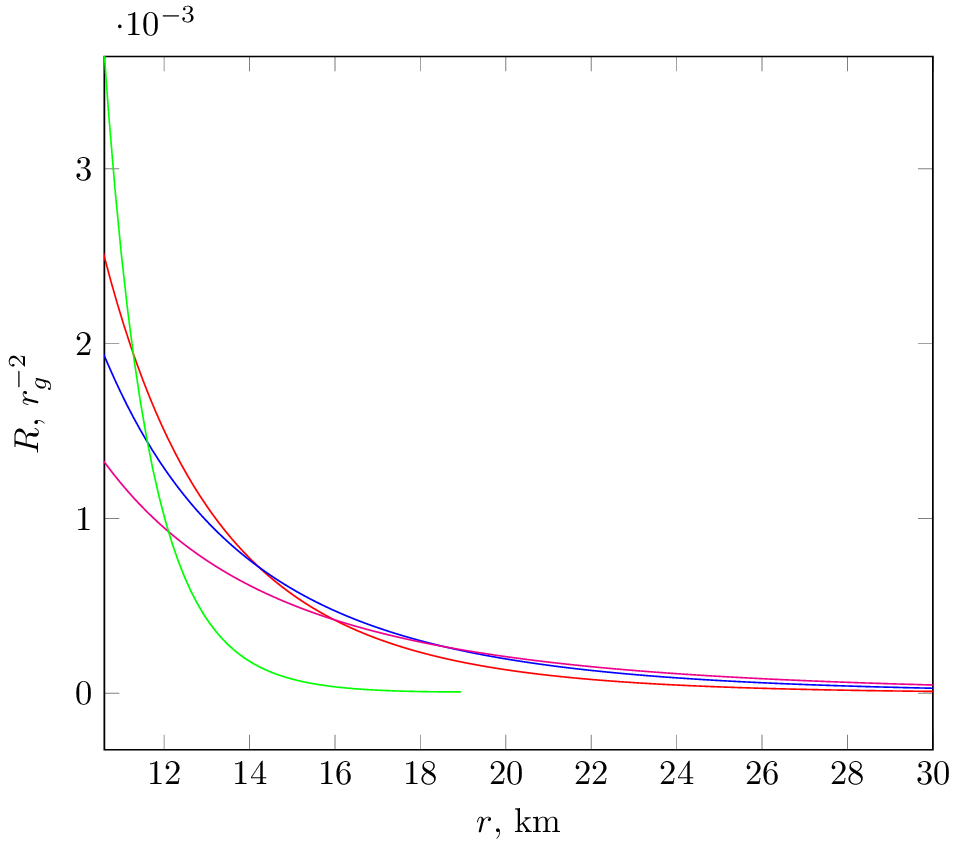}\includegraphics[scale=0.7]{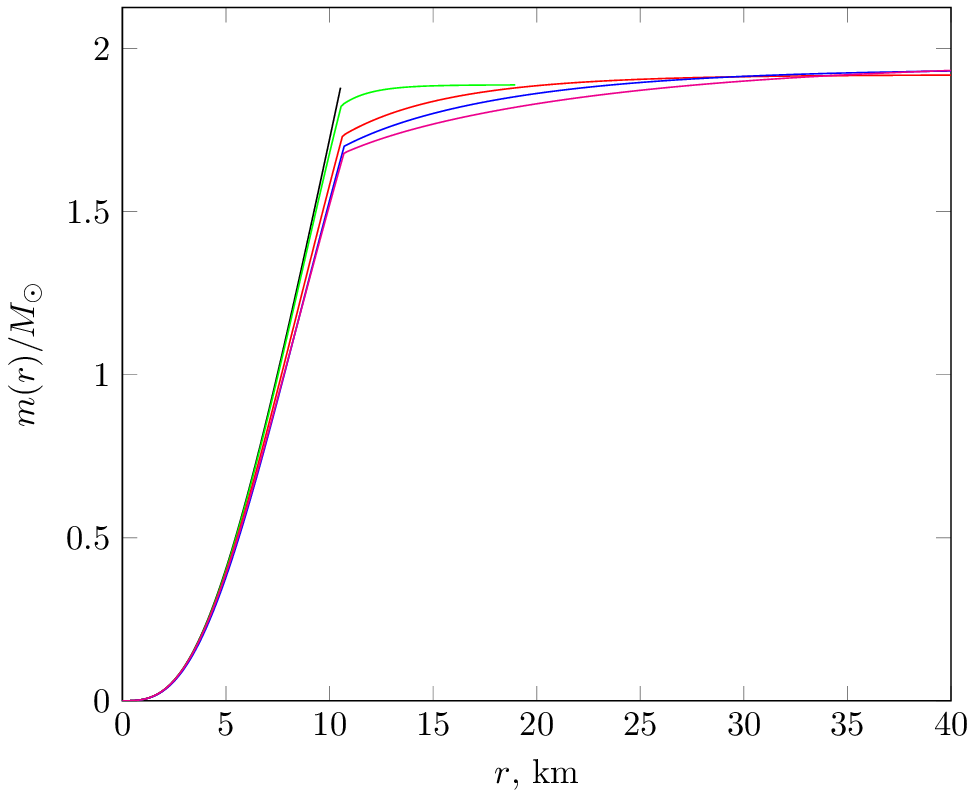}
\caption{Left panel: The dependence of scalar curvature outside of
star for stellar configuration with maximal mass for simple quark
EoS. Right panel: the stellar mass profile $m(r)$ for stellar
configurations with maximal gravitational mass.} \label{Fig11}
\end{center}
\end{figure*}

On the other hand, Figs. \ref{Fig3}, \ref{Fig5}, \ref{Fig7},
\ref{Fig9} and \ref{Fig11} represent the Ricci scalar curvature
evolution outside the star. For $R^2$ gravity, the curvature turns
out to be positive on star surface and decreases outside the star
so that $R\rightarrow 0$ at $r\rightarrow\infty$.
In the aforementioned figures, one can see the following features: Firstly the
maximal gravitational mass $M_{max}$ (defined as ${\rm d}M/{\rm d}r_s=0$) of star for
each EoS slowly increases with 
$\alpha$. The value $\Delta M=M_{max}(\alpha)-M^{\rm GR}_{max}\sim
\sqrt{\alpha}$. At first glance one cannot discriminate between
$R^2$-gravity (for the considered values of $\alpha$) and GR
because such mass differences lie within typical errors of
neutron star mass measurements from pulsar timing. Indeed,
from Table I in \cite{Lattimer2} one can see that
1-$\sigma$-uncertainties for neutron star masses for X/ray-optical binaries
and neutron star-white dwarf binaries are $\sim 0.1-0.2M_\odot$
with few exceptions. Only for neutron stars with small masses precise measurements of masses with errors $<0.01M_\odot$
(neutron star-neutron star binaries) are available. In conclusion, we have no well-established data
on radii of these low-mass neutron stars.

Another in principle observable astrophysical quantity for these gravitational configurations is
the so-called the surface gravitational redshift $z_{s}$
\cite{Cottam}, which is defined as
\begin{equation}
z_{s}=(1-2M_{s}/r_{s})^{-1/2}-1\,,
\end{equation}
and therefore $z_S$ is determined by stellar mass $M_s$.
According to our results, in the context of $R^2$-gravity for
$\alpha>0$, we conclude that $z_{s}<z_{s,\, {\rm GR}}$. Moreover,
the expected absolute deviation of the surface redshift from the
GR value for a given mass, i.e., $\Delta z=z_{s, \,{\rm
GR}}-z_{s}$ turns out to be weakly dependent on the gravitational
mass for typical masses of neutron stars. In principle this
discrepancy could help to discriminate between GR and
$R^2$-gravity, and by extension to establish the validity of
different classes of $f(R)$ models and other extended theories of
gravity. Indeed, in the case of $R^2$-gravity we have shown that
$M_s<M$. Thus, the measure of the gravitational redshift would allow us to label the star by its
stellar mass. On the other hand, as stated above, by means of binary systems one can measure
the gravitational mass $M$. Accordingly, the discrepancy between these two quantities could in principle witnesses in favor of $R^2$-gravity, or eventually other extended theories of gravity for which
$M\neq M_s$.  Unfortunately, at the present moment,
one have no well-defined data about neutron-star gravitational redshifts nor radii.

Another interesting feature is the increase of star radius for a given
gravitational mass close to maximal mass $M_{max}$ when compared to GR counterparts (see Table \ref{Table1}). In principle
this fact may open the possibility to obtain an upper limit on parameter
$\alpha$ by using observational data for neutron stars radii. Unfortunately, at
the present time we have no well-established such data.
\\

From the gravitational mass-radius relation
in Figs. \ref{Fig2}, \ref{Fig4}, \ref{Fig6}, \ref{Fig8} and \ref{Fig10}
we see that for purely hadron EoS, the radii of neutron stars with
typical masses $\sim 1.4-1.5 M_\odot$ hardly 
depends on the value of $\alpha$. This effect is more
noticeable for SLy and MPA1 EoS.
As an illustrative example, let us mention that for GM1 EoS with inclusion of hyperons we see in Fig \ref{Fig8} that for several values of $\alpha$ the mass-radius
curves intersect in the vicinity of $1.6M_\odot$. Therefore the data
corresponding to the mass-radius relation in the vicinity of either large or small masses
can help to discriminate between GR and the gravitational modification under study here.
Similar statements can be made for the other two neutron stars EoS.
\\

Finally, for quark stars we have the following
scenario: the density on star surface of strange stars is nonzero,
namely $\rho_{s}=4B$. Therefore, the scalar curvature evaluated at
the star's surface, as predicted by GR, would be $R_{s}=32\pi B$
whereas outside the star, the scalar curvature drops to zero
suddenly. On the contrary, in $R^2$-gravity one can construct
solution with smooth decreasing of scalar curvature as can be seen
in Fig. \ref{Fig11}.

\begin{table}[htbp]
\begin{tabular}{|c|c|c|c|c|c|}
  \hline
  EoS  & $\alpha$,         & $M_{max}$,    & $M_{s,\, max}$, & $\Delta M_{max},$ & $\Delta r_{max}$, \\
       & $10^{10}$ cm$^2$    & $M_{\odot}$   & $M_{\odot}$     &  $M_{\odot}$      &        km         \\
  \hline
        & 0  & 2.23 & 2.23 & 0 &  0 \\
        & 0.5 & 2.22 & 2.18 & 0.045 (1.94) & 0.05   \\
  AP4   & 5  & 2.24 & 2.12 & 0.182 (1.82) & 0.45   \\
        & 10  & 2.26 & 2.12 & 0.232 (1.88) &  0.63  \\
        & 20  & 2.29 & 2.11 & 0.296 (1.91) & 0.84   \\
        \hline
        & 0  & 2.05 & 2.05  & 0    & 0   \\
        & 0.5 & 2.05 & 2.02 & 0.042 (1.92) &  0.08   \\
  SLY   & 5  & 2.10 & 1.97  & 0.176 (1.74) & 0.58  \\
        & 10  & 2.13 & 1.95 & 0.237(1.84) & 0.80   \\
        & 20  & 2.16 & 1.94 & 0.294 (1.86) & 0.96   \\
        \hline
        & 0  & 2.49 & 2.49 & 0 & 0    \\
        & 0.5  & 2.49 & 2.45 & 0.045 (2.29) & 0.045    \\
  MPA1  & 5  & 2.51 & 2.38 & 0.188 (1.99) & 0.55   \\
        & 10  & 2.54 & 2.37 & 0.256 (2.01) & 0.78   \\
        & 20  & 2.57 & 2.34 & 0.320 (2.02) &  1.0  \\
  \hline
        & 0  & 1.93 & 1.93  & 0 & 0  \\
        & 0.5  & 1.93 & 1.90 & 0.035(1.88) & 0.06   \\
  GM1nph & 5 & 1.96 & 1.80   & 0.169 (1.85) &  0.32 \\
        & 10  & 1.98 & 1.76  & 0.237 (1.89) &  0.47 \\
        & 20  & 2.01 & 1.73  & 0.296 (1.82) &  0.60  \\
        \hline
        & 0  & 1.88 & 1.88  & 0 & 0  \\
        & 0.5  & 1.89 & 1.82 & 0.072 (1.72) &  0.075   \\
  QEoS   & 5  & 1.92 & 1.73 & 0.209 (1.71) & 0.31   \\
        & 10  & 1.94 & 1.70   & 0.264 (1.71) & 0.41 \\
        & 20  & 1.96 & 1.67  & 0.313 (1.82) & 0.51  \\
        \hline
\end{tabular}
\caption{Parameters of compact stars for various equations of
state in General Relativity (i.e., $\alpha=0$) and for several
values of $\alpha$ in $f(R)=R+\alpha R^2$ gravity theories. The
quantity $\Delta M_{max}$ holds for the maximal contribution of
the gravitational sphere outside the star in the total value of
the gravitational mass (having included the stellar mass).
On the fourth column, the value for the corresponding gravitational mass is
given in brackets. In the last column we show the 
maximum difference in the star radius as compared with the General Relativity counterpart. This maximum difference 
takes place for a star with a mass
equal to the maximal mass in General Relativity for a given equation of state.}
\label{Table1}
\end{table}

\section{Results for $\alpha<0$}
\label{Sec:V}
In this Section we have considered the case $\alpha=-0.05$ km$^2$ as an illustrative
example in order to get a better understanding of the behaviour of the system of Eqs. \eqref{5}-\eqref{8} when applied to $R^2$-gravity models with negative parameter $\alpha$.
The considered central densities 
for the studied examples below correspond to stars possessing a GR maximal mass for each of the
chosen EoS. The system \eqref{5}-\eqref{8}  is thus integrated up to a large enough
distance ($r\approx 3\cdot10^5$ km).

We have analysed the $B(r)$ dependence with the radial coordinate
and studied the behaviour of this metric function as
$r\rightarrow\infty$ so by the use the Eq. (\ref{mass2}) we have
tried to extract the corresponding gravitational mass. Another
(but completely equivalent) way we have followed consisted of
trying to find such an initial value $B(0)$ for which the function
 ${B}(r)\rightarrow 1$ as $r\rightarrow\infty$. We shall
designate this value as $\bar{B}(0)$. For the latter method,  one
can consider a sequence of $\bar{B}_{i}(0)$ values such that
${B}(r_{i})=1$. Then extrapolating this sequence for
$r_{i}\rightarrow\infty$ gives the required $\bar{B}(0)$. The
gravitational mass can then be calculated according to Eq.
(\ref{mass1}).

We have also performed the following calculations: after having
assumed that $m(r)$ varies slowly at large distances within some
interval $\Delta r$ (we take $\Delta r\approx 10^4$ km), we have
approximated $B(r)$ by a functional dependence
\begin{eqnarray}
B(r)=B_{\infty}\left(1-\frac{2m}{r}\right)
\end{eqnarray}
in this  $\Delta r$ interval. Thus, as a rough estimation one can
identify the corresponding mass for given $r$ with $m$.
For GR this scheme gives correct value for gravitational mass,
whereas for $R^2$ gravity our method shows that that the found
mass grows with the radial distance.

In order to illustrate the reasoning, we have chosen $B(0)=0.1$. Table \ref{Table2} below summarises the found results for the MPA1 EoS
case.
 We have also calculated
the values $\bar{{B}}_{i}(0)$ at which for $r_{i}$ the condition
$1-{B}(r_{i})=\beta$, is accomplished with $\beta=10^{-4}$.
In GR $B(r)$ exactly scales like $1/r$ and therefore one can
extract the gravitational mass $M$ using dependence $B(r)$ for an
arbitrary $B(0)$ value, by simply approximating this dependence to
the well-known function $B_{\infty}(1-2M/r)$. Thus, for
viable theories beyond GR the function $m(r)$
 is expected to tend to constant value asymptotically, i.e., there is a plateau for
$m(r)$.\\

In the context of $f(R)=R+\alpha R^2$ ($\alpha<0$) models, one can
assume that the aforementioned plateau for the function $m(r)$ will be
found at sufficiently large distances. Nonetheless, our method
shows that the approximation of $\bar{B}(0)$ (or equivalently
$B_{\infty}$ for an arbitrary $B(0)$) strongly depends on the considered
interval of distances.

In particular we have considered the approximation of $B(r)$
by a function of the form $B_\infty(1-2m/r)$ in intervals
$r_{i}<r<r_{i}+10^4$ km for some $r_{i}$ and results have been
shown in Table \ref{Table3}). In this case the dependence
$(1-B(r)/B_\infty)r/2$ has a maximum near $r_{i}$.
We have also considered the interval $10^3\, {\rm km} < r
<5\cdot10^4$ km for $B(r)$. In this case we have for
$B_\infty=1.488271$ and the maximal value for ${m}(r)$ is
$2.71M_\odot$. If for instance the consider distance interval is
now $2\cdot10^4\, {\rm km}<r<7\cdot10^4$ km one obtains
$B_\infty=1.488362$ and maximal of $m(r)$ is $3.15M_\odot$. Thus
our method leads us to conclude that $B_\infty-B(r)\nsim r^{-1}$
even for large distances.
Consequently, the main point of our analysis is that the function
$B_\infty-B(r)$ (or $1-B(r)$ for calibrated initial condition for
$B(0)$) tends to $0$ more slowly than $r^{-1}$ at
$r\rightarrow\infty$. Therefore we are led to conclude that we
have no well-defined limit of mass since the function $m(r)$ seems
to grow with radial distance. In fact the value of $\bar{B}(0)$
(or $B_{\infty}$) decreases (increases) with growing of maximal
limit of radial interval under consideration.
\\

Of course one can assume that the results above only take place
for some EoS only. In order to see if this is true,  we have
considered all the other three-nucleon EoS from Sec.
\ref{Sec:III}, namely AP4, SLy and GM1. 
For those, we have obtained
that the $m(r)$ growth rate with the radial coordinate slightly depends on the specific
EoS, but there is no qualitative difference with respect to the
trend fully explained above.

\begin{table}
\begin{centering}
\begin{tabular}{|c|c|c|c|c|c|}
  \hline
  $r_{i}$, $10^4$ km  & $B_{i}$  & $\bar{B}_{i}(0)$ & $m(r_{i})$ & $B_{i,\,{\rm GR}}$\\
  \hline
  1   & 1.487088 & 0.067178 & 2.64  & 1.482622    \\
  2   & 1.487687 & 0.067151 & 2.83  & 1.483165    \\
  3   & 1.487900 & 0.067141 & 3.02  & 1.483346    \\
  4   & 1.488013 & 0.067136 & 3.19  & 1.483437    \\
  5   & 1.488085 & 0.067133 & 3.37  & 1.483491    \\
  6   & 1.488135  & 0.067131 & 3.55  & 1.483528   \\
  7   & 1.488173  & 0.067129 & 3.72  & 1.483553   \\
  8   & 1.488203  & 0.067128 & 3.90  & 1.483573   \\
  9   & 1.488227  & 0.067127 & 4.07  & 1.483588  \\
  10  & 1.488248  & 0.067126 & 4.25  & 1.483600   \\
  12  & 1.488280  & 0.067124 & 4.60  & 1.483618   \\
  14  & 1.488305 & 0.067123 & 4.95  & 1.483631    \\
  \hline
\end{tabular}
\caption{Calculation of $B(r_{i})\equiv B_i$ at some $r_{i}$ for
$B(0)=0.1$ and MPA1 equation of state. The fourth column shows $m(r_{i})$. Radial
distances have been considered in the interval $r_i-5\cdot 10^3\, {\rm
km}<r<r_i+5\cdot10^3$ km. As observed, masses $m(r_i)$ grow with $r_i$.
As expected, numerical calculations in General Relativity give the
well-known result, i.e., the function $(B_\infty-B(r))r/2$ remains
constant being the gravitational mass is $2.48M_\odot$  for this case.}
\label{Table2}
\end{centering}
\end{table}

\begin{table}[H]
\begin{centering}
\begin{tabular}{|c|c|c|c|}
  \hline
  $r_{i}$, $10^4$ km  & $B_{\infty}$  & $\bar{B}(0)$ & ${m}_{max}$  \\
  \hline
  1   & 1.488284 & 0.067192 & 2.74      \\
  2   & 1.488325 & 0.067190 & 2.92      \\
  3   & 1.488352 & 0.067189 & 3.05      \\
  4   & 1.488372 & 0.067188 & 3.27      \\
  5   & 1.488387 & 0.067187 & 3.44       \\
  6   & 1.488400 & 0.067186 & 3.62       \\
  $0.1<r<5$ & 1.488271 & 0.067919 & 2.71 \\
  $2<r<7$ & 1.488362 & 0.067188 & 3.18 \\
  \hline
\end{tabular}
\caption{Results for $B_{\infty}$ in the interval
$r_{i}<r<r_{i}+10^4$ km for $B(0)=0.1$ for the MPA1 equation of state
and corresponding of $\bar{B}(0)$ in same interval. We give the
maximum value of ${m}(r)$ (${m}_{max}$) for corresponding the
$\bar{B}$ and $B_{\infty}$.} \label{Table3}
\end{centering}
\end{table}

\section{Conclusions}
\label{Sec:VI}

In this paper we have investigated the main features and existence
of realistic models of compact neutron and quarks stars in one
paradigmatic extension of General Relativity. Namely we have
studied these static and spherically symmetric configurations for
a class of fourth-order $f(R)$ extended theories of
gravity, whose gravitational Lagrangian adopts the form
$f(R)=R+\alpha R^2$. As explained in the bulk of the article, the
resolution of the generalised system of Tolman-Oppenheimer-Volkoff
equations for positive $\alpha$ has indeed provided the existence
of solutions, whereas for negative $\alpha$ it has been impossible
to recover a weak-field (Newtonian) limit far away from the star,
concluding the impossibility of stable configurations in such a
region of the parameter space.
\\

Note that our analysis has not considered any perturbative
approximation whatsoever when dealing with the system of equations resolution nor a shooting method has been required owing to the chosen variables. Thus, we have found that for the same
central density the stellar mass as bounded by the star surface
decreases when the parameter $\alpha$ increases. This result can be
understood as a proof that in extended theories of gravity,  a more intense gravitational
interaction leads to stable stars which for the same mass in Einsteinian gravity would either not exist or be condemned to
be black holes. 
More specifically, for the parameters space of quadratic $f(R)$ models which are in agreement with theoretical constraints and cosmological data, the $f(R)$ predictions for the neutron stars maximal mass ranged from 2-2.6\,$M_{\odot}$ (see Table \ref{Table1})
 and therefore agreed with observed stars as those found in \cite{others_masses}. Similar results were found for the so-called Hu-Sawicki $f(R)$ model 
 \cite{Resco:2016upv}.\\
Also, we have seen that, unlike
previous works \cite{Capo} which
naively defined the mass as the matter integral over the star volume,
we have shown how a more careful matching of interior and exterior
regions causes the total gravitational mass to be unbounded for
sufficiently large values of $\alpha$ (with specific values
depending on the equation of state under consideration).
Consequently this more general consideration of conditions to be
imposed at the edge of the star is twofold.
\\

First, these reasonable matching conditions imply that  the scalar
curvature is not identically null on the edge of the star, even so
(for $\alpha>0$) it is exponentially suppressed and goes to zero and accordingly
asymptotic flatness is eventually recovered. This can be thought
of as being equivalent to the existence of an additional energetic
content around the star which might be assigned to the additional
extra scalar mode present in $f(R)$ theories. It is precisely this
mode which can help to prevent the gravitational collapse
and increase the gravitational mass as observed by a distant
observer. Thus in the event of merging objects of this kind, the
available energy would be higher than General Relativity
counterparts and the emission of gravitational waves from such a
configuration might be feasible and detectable by present and
future interferometers \cite{Resco:2016upv,Damour:1996ke, Berti:2015itd,Koyama:2011xz,Sibandze:2016agp, BeltranJimenez:2017doy}.
\\

Secondly, we have indeed corroborated the
existence of exterior static and spherically symmetric physical solutions differing from the usual Schwarzschild solution (non Ricci flat) provided $f(R)$ theories are assumed to represent the underlying gravitational theory in high-curvature environments.
This fact constitutes a straightforward violation of the Jebsen-Birkhoff theorem  in $f(R)$ theories \cite{Clifton:2012ry,Nzioki:2013lca}. 
Whence it seems that the most natural spacetime solutions as generated by a spherical stable neutron or quark star is not the Schwarzschild solution once Einsteinian gravity is replaced by simple extensions of the standard Einstein-Hilbert Lagrangian.
\\

Concerning the observational imprints to be expected from our
results, let us claim that the net effect on the total mass is a
faint, although $\alpha$-parameter, dependent increase in the mass
as perceived by a distant observer with respect to the General
Relativity prediction for the same central density and equation of
state. Also, we have concluded the increase of star radii in
vicinity of maximal masses as well as the radii reduction for
stars with masses of the order of the Sun mass.

Then, we can think of at least two ways for discriminating between
Einsteinian gravity and possible modifications
from scalar-tensor theories, 
 and by extension other extended theories of gravity with analogous effects. Firstly, high-accuracy measurements of surface redshift
can shed some light on the underlying theory. Indeed, for the same asymptotic gravitational mass, the pure stellar mass
bounded by surface as generated by $R^2$-gravity would be smaller than the Einsteinian value and
therefore the measured surface redshift 
would be smaller than predicted.
Secondly, another observable effect would be the determination of gravitational maximal masses, which we have calculated for competitive both neutron plasma and quark stars, using
X/ray-optical binaries and neutron star-white dwarf binaries.
Notwithstanding, the present precision in mass measurements is not enough to discriminate between competing theories yet.
Both methods have been shown here as competitive proposals to break the degeneracy problem afflicting alternative theories of gravity.\\

Finally, for $\alpha<0$ our analysis has shown that outside the star the additional energetic effect explained above, i.e., the
appearance of a {\it gravitational sphere}, is also present. Nevertheless, the scalar curvature suffers
damped oscillations instead of an exponential attenuation. Moreover, in these cases the
contribution of the {\it gravitational sphere} into the gravitational mass
has been shown to increase indefinitely with radial distance. Therefore we conclude that
$R^2$-gravity with $\alpha<0$ predictions are not consistent
with the existence of realistic models of compact relativistic stars, at least for the studied equations of state.
The inadmissibility of $R^2$ gravity with $\alpha<0$ was made
earlier in cosmological contexts so our result shows that these classes of models would not be acceptable in the astrophysical context either in order to
 obtained finite asymptotic masses.
\\

As a final comment  let us mention that,  the value for $\alpha$ in the Starobinsky inflationary model in good agreement with Planck latest results 
provides a much smaller value \cite{Riotto} than those considered in this article.  
%
Thus, the $R^2$ models presented here should be understood as effective corrections to General Relativity appearing at curvatures of the order of those 
characterising the dynamics of neutron/quark stars.\\

\begin{acknowledgements}
We would like to thank Miguel Aparicio Resco for useful and meaningful discussions on the topic.
The work of AVA is supported by project {\color{blue}
1.4539.2017/8.9} (MES, Russia).
SDO  is supported by MINECO (Spain), project FIS2016-76363-P and
by CSIC I-LINK1019 Project.
AdlCD acknowledges financial support from Consolider-Ingenio
MULTIDARK CSD2009-00064,  i-Link1019, Spanish Ministry of Economy
and Science and FIS2016-78859-P, European Regional Development
Fund and Spanish Research Agency (AEI), Management Committee
Member for the Cosmology group - University of Cape Town, CANTATA
COST action CA15117,  EU Framework Programme Horizon 2020,
 University of Cape Town Launching Grants Programme and National Research Foundation grant 99077 2016-2018, Ref. No. CSUR150628121624 and NRF Incentive Funding for Rated Researchers (IPRR), Ref. No. IFR170131220846.

\end{acknowledgements}

\begin{thebibliography}{99}

\bibitem{Riess} A.~G.~Riess {\it et al.}  [Supernova Search Team Collaboration], Astron.\ J.\  {\bf 116}, 1009 (1998) [arXiv:astro-ph/9805201]; A.~G.~Riess {\it et al.}  [Supernova Search Team Collaboration],
Astrophys.\ J.\  {\bf 607}, 665 (2004)[arXiv:astro-ph/0402512].

\bibitem{Perlmutter} S.~Perlmutter {\it et al.}  [Supernova Cosmology Project Collaboration], Astrophys.\ J.\  {\bf 517}, 565 (1999) [arXiv:astro-ph/9812133].

\bibitem{Capozziello2} S. Capozziello, S. Carloni, A. Troisi, Recent Res. Dev. Astron. Astrophys. {\bf 1}, 625 (2003).

\bibitem{Odintsov1} S. Nojiri, S.D. Odintsov, Phys. Rev. D {\bf 68}, 123512 (2003); Phys. Lett. B  {\bf 576}, 5 (2003).

\bibitem{Turner} S.~M.~Carroll, V.~Duvvuri, M.~Trodden and M.~S.~Turner, Phys.\ Rev.\ D. {\bf 70},  043528 (2004).


\bibitem{Odintsov-3} S. Nojiri and S. D. Odintsov, Phys. Rept. {\bf 505}, 59 (2011) [arXiv:1011.0544 [gr-qc]]; eConf C 0602061, 06 (2006) [Int. J. Geom. Meth. Mod. Phys. {\bf 4}, 115 (2007)] [hep-th/0601213];
arXiv:1306.4426 [gr-qc].

\bibitem{Capozziello_book} S. Capozziello and V. Faraoni, \textit{Beyond Einstein Gravity} (Springer) New York  (2010).

\bibitem{Capozziello4} S. Capozziello and M. De Laurentis, Phys. Rept.{\bf  509}, 167 (2011) [arXiv:1108.6266[gr-qc]].



\bibitem{Cruz-Cosmography} 
  A.~de la Cruz-Dombriz, P.~K.~S.~Dunsby, O.~Luongo and L.~Reverberi,   
  JCAP {\bf 1612} (2016) no.12,  042 
  [arXiv:1608.03746 [gr-qc]].

  \bibitem{Bamba:2012}
  K.~Bamba, S.~Capozziello, S.~Nojiri and S.~D.~Odintsov,
  Astrophys.\ Space Sci.\  {\bf 342}, 155 (2012) [arXiv:1205.3421 [gr-qc]].

\bibitem{delaCruzDombriz:2012xy}
  A.~de la Cruz-Dombriz and D.~Saez-Gomez,   
  Entropy {\bf 14}, 1717 (2012) 
  [arXiv:1207.2663 [gr-qc]].

\bibitem{Oppenheimer:1939ne}
 J.~R.~Oppenheimer and G.~M.~Volkoff,
  Phys.\ Rev.\  {\bf 55}, 374 (1939);
  R.~C.~Tolman,
  Phys.\ Rev.\  {\bf 55}, 364 (1939).



\bibitem{Dobado:2011gd}
  A.~Dobado, F.~J.~Llanes-Estrada and J.~A.~Oller,
  Phys.\ Rev.\ C {\bf 85}, 012801 (2012)
  [arXiv:1107.5707 [gr-qc]].



\bibitem{Demorest:2010bx}
  P.~Demorest, T.~Pennucci, S.~Ransom, M.~Roberts and J.~Hessels,   
  Nature {\bf 467}, 1081 (2010)   [arXiv:1010.5788 [astro-ph.HE]].

\bibitem{Antoniadis:2013pzd}
  J.~Antoniadis {\it et al.},
  Science {\bf 340} (2013) 6131
  doi:10.1126/science.1233232
  [arXiv:1304.6875 [astro-ph.HE]].

\bibitem{others_masses}
  O.~Barziv, L.~Kaper, M.~H.~van Kerkwijk, J.~H.~Telting and J.~van Paradijs,   
  Astron.\ Astrophys.\  {\bf 377}, 925 (2001)     [astro-ph/0108237];
  M.~L.~Rawls, J.~A.~Orosz, J.~E.~McClintock, M.~A.~P.~Torres, C.~D.~Bailyn and M.~M.~Buxton, 
  Astrophys.\ J.\  {\bf 730} (2011) 25   [arXiv:1101.2465 [astro-ph.SR]]; 
  T.~Munoz-Darias, J.~Casares and I.~G.~Martinez-Pais,   
  Astrophys.\ J.\  {\bf 635} (2005) 502   [astro-ph/0508547];  
  D.~J.~Nice, E.~M.~Splaver, I.~H.~Stairs, O.~Loehmer, A.~Jessner, M.~Kramer, 2 and J.~M.~Cordes, 
  Astrophys.\ J.\  {\bf 634}, 1242 (2005) [astro-ph/0508050]. 

\bibitem{Dexheimer:2007mt}
  V.~A.~Dexheimer, C.~A.~Z.~Vasconcellos and B.~E.~J.~Bodmann,   
  Phys.\ Rev.\ C {\bf 77} (2008) 065803  
  [arXiv:0708.0131 [astro-ph]].

  
\bibitem{Astashenok-1} A. Astashenok, S. Capozziello, S. Odintsov, JCAP \textbf{12}, 040 (2013) [arXiv:1309.1978 [gr-qc]];

Phys. Rev. D {\bf 89}, 103509 (2014)  [arXiv:1401.4546 [gr-qc]];

Astrophys. Space Sci. 355, 333 (2015) [arXiv:1405.6663 [gr-qc]];

JCAP 1501, 001 (2015) [arXiv:1408.3856 [gr-qc]].



\bibitem{Jing:2015ota}
  Z.~Jing, D.~Wen and X.~Zhang,
  Sci.\ China Phys.\ Mech.\ Astron.\  {\bf 58}, no. 10, 109501 (2015).
  doi:10.1007/s11433-015-5694-3


\bibitem{Kobayashi-Maeda} T.~Kobayashi and K.~i.~Maeda, Phys.\ Rev.\ D {\bf  78}, 064019 (2008) [arXiv:0807.2503 [astro-ph]].

\bibitem{Upadhye-Hu} A.~Upadhye and W.~Hu, Phys.\ Rev.\ D {\bf  80}, 064002 (2009) [arXiv:0905.4055 [astro-ph.CO]].


\bibitem{Feng:2017hje}
  W.~X.~Feng, C.~Q.~Geng, W.~F.~Kao and L.~W.~Luo,
  arXiv:1702.05936 [gr-qc].


\bibitem{Pannia:2016qbj}
  F.~A.~Teppa Pannia, F.~GarcAa, S.~E.~Perez Bergliaffa, M.~Orellana and G.~E.~Romero,
  Gen.\ Rel.\ Grav.\  {\bf 49}, 25 (2017) arXiv:1607.03508 [gr-qc].

\bibitem{Wojnar:2016bzk}
  A.~Wojnar and H.~Velten,
  Eur.\ Phys.\ J.\ C {\bf 76}, 697 (2016) arXiv:1604.04257 [gr-qc].

\bibitem{Arapoglu:2016ozr}
  S. Arapo\v{g}lu, S. \c{C}ikinto\v{g}lu, K. Yavuz Ek\c{s}i,
  arXiv:1604.02328 [gr-qc].

\bibitem{Katsuragawa:2015lbl}
  T.~Katsuragawa, S.~Nojiri, S.~D.~Odintsov and M.~Yamazaki,
  Phys.\ Rev.\ D {\bf 93}, 124013 (2016) , arXiv:1512.00660 [gr-qc].



\bibitem{Fiziev:2015xpa}
  P.~P.~Fiziev,
  arXiv:1506.08585 [gr-qc].

\bibitem{Hendi:2015pua}
  S.~H.~Hendi, G.~H.~Bordbar, B.~Eslam Panah and M.~Najafi,
  Astrophys.\ Space Sci.\  {\bf 358}, 30 (2015), arXiv:1503.01011 [gr-qc].

\bibitem{Momeni:2015vwa}
  D.~Momeni, H.~Gholizade, M.~Raza and R.~Myrzakulov,
  Int.\ J.\ Mod.\ Phys.\ A {\bf 30}, 1550093 (2015) arXiv:1502.05000 [gr-qc].

\bibitem{Zubair:2016kov}
  M.~Zubair and G.~Abbas,
  Astrophys.\ Space Sci.\  {\bf 361}, 342 (2016).

\bibitem{Bakirova:2016ffk}
  E.~Bakirova and V.~Folomeev,
  Gen.\ Rel.\ Grav.\  {\bf 48}, 135 (2016) Erratum: [Gen.\ Rel.\ Grav.\  {\bf 48}, 164
  arXiv:1603.01936 [gr-qc].

\bibitem{Resco:2016upv}
  M.~Aparicio Resco, \'{A}.~de la Cruz-Dombriz, F.~J.~Llanes Estrada and V.~Zapatero Castrillo,
  Phys.\ Dark Univ.\  {\bf 13}, 147 (2016) arXiv:1602.03880 [gr-qc].

\bibitem{Moraes:2015uxq}
  P.~H.~R.~S.~Moraes, J.~D.~V.~Arba\~{n}il and M.~Malheiro,
  JCAP {\bf 1606}, 005 (2016) arXiv:1511.06282 [gr-qc].

\bibitem{Sharif:2015jaa}
  M.~Sharif and Z.~Yousaf,
  Can.\ J.\ Phys.\  {\bf 93}, 905 (2015).

\bibitem{Sotani:2017pfj}
  H.~Sotani and K.~D.~Kokkotas,
  Phys.\ Rev.\ D {\bf 95}, 044032 (2017) arXiv:1702.00874 [gr-qc].

\bibitem{Laurentis} S.~Capozziello, M.~De Laurentis, I.~De Martino, M.~Formisano and S.~D.~Odintsov, Phys.\ Rev.\ D {\bf 85}, 044022 (2012) [arXiv:1112.0761 [gr-qc]].

\bibitem{Laurentis2} S.~Capozziello, M.~De Laurentis, S.D.~Odintsov and A.~Stabile, Phys.\ Rev.\ D {\bf 83}, 064004 (2011) [arXiv:1101.0219 [gr-qc]].



  \bibitem{Arapoglu} S. Arapo\v{g}lu, C. Deliduman, K. Yavuz Ek\c{s}i, JCAP {\bf 1107}, 020 (2011) arXiv:1003.3179v3[gr-qc].

\bibitem{Alavirad} H. Alavirad, J.M. Weller, arXiv:1307.7977v1[gr-qc].


\bibitem{Yazadjiev2014} S.S. Yazadjiev, D.D. Doneva,  K.D. Kokkotas, K.V. Staykov JCAP 1406, 003 (2014);
S.S. Yazadjiev, D.D. Doneva,  K.D. Kokkotas  Phys. Rev. D 91,
084018 (2015).

\bibitem{Doneva2014a} D.D. Doneva, S.S. Yazadjiev,  K.V. Staykov, K.D. Kokkotas, Phys.
Rev. D 90, 104021 (2014) arXiv:1408.1641 [gr-qc]; Phys. Rev. D 92, 064015 (2015).

\bibitem{Henttunen} K. Henttunen, I. Vilja, JCAP 1505, 001 (2015)
[arXiv:1408.6035[gr-qc]];

K. Henttunen, I. Vilja, Phys.Lett. B 731, 110 (2014)
arXiv:1110.6711 [astro-ph.CO]].

\bibitem{Astashenok2015} A. Astashenok, S. Capozziello, S. Odintsov, Phys. Lett. B \textbf{742}, 160 (2015).

 \bibitem{Clifton:2012ry}
  T.~Clifton, P.~Dunsby, R.~Goswami and A.~M.~Nzioki,
  Phys.\ Rev.\ D {\bf 87}, no. 6, 063517 (2013)
  [arXiv:1210.0730 [gr-qc]].

\bibitem{Fiziev} P. Fiziev, K. Marinov, Bulgarian Astronomical Journal, 23, 3,
(2015) arXiv:1412.3015 [gr-qc].

\bibitem{SLy} Chabanat, E. and Bonche, P. and Haensel, P. and Meyer,
J. and Schaeffer, R., Nucl. Phys. A635, 231 (1998).

\bibitem{SLy-4} Douchin, F. and Haensel, P., Astr. Ap. 380, 151 (2001).

\bibitem{APR} Akmal, A. and Pandharipande, V.~R. and Ravenhall, D.~G.,
Phys. Rev. C 58, 1804 (1998).

\bibitem{GM} Glendenning, N.~K. and Moszkowski, S.~A., Phys. Rev. Lett.
67, 2414 (1991).

\bibitem{GM-1} Oertel, M. and Provid{\^e}ncia, C. and Gulminelli, F. and
Raduta, A.~R., J. Phys. G Nucl. Phys. 42, 075202 (2015).


\bibitem{MPA} M{\"u}ther, H. and Prakash, M. and Ainsworth, T.~L.,
Phys. Lett. B199, 469 (1987).

\bibitem{Camenzind} M. Camenzind, \textit{Compact Objects in Astrophysics}, Springer (2007).

\bibitem{Eksi} {C. G\"{u}ng\"{o}r, K.Y. Ek\c{s}i, Proceedings of conference
``Advances in Computational Astrophysics: methods, tools and
outcomes'' (Cefalu (Sicily, Italy), June 13-17, 2011),
[arXiv:1108.2166v2  [astro-ph.SR]].}

\bibitem{Itoh} N. Itoh, Progress of Theoretical Physics 44, 291 (1970).

\bibitem{Witten} E.Witten, Phys. Rev. D 30, 272 (1984).

\bibitem{Jaffe} R.L. Jaffe, F. E. Low, Phys. Rev. D. 19, 2105
(1979).

\bibitem{Simonov} Yu. Simonov, Phys. Lett. B. 107 (1981).

\bibitem{Sterg} N. Stergioulas, Living Rev. Rel. 6, 3 (2003) arXiv:gr-qc/0302034 [gr-qc].

\bibitem{Lattimer2} J.M. Lattimer, Annu. Rev. Nucl. Part. Sci. 62, 485
(2012).

\bibitem{Cottam} J. Cottam, F. Paerels, M. Mendez, Nature 420, 51
(2002).









\bibitem{Capo} S. Capozziello, M. De Laurentis, R. Farinelli and S. D. Odintsov, arXiv:1509.04163 [gr-qc].


\bibitem{Damour:1996ke}
  T.~Damour and G.~Esposito-Farese,   
  Phys.\ Rev.\ D {\bf 54}, 1474 (1996) 
  [gr-qc/9602056].


\bibitem{Berti:2015itd}
  E.~Berti {\it et al.},   
  Class.\ Quant.\ Grav.\  {\bf 32} (2015) 243001 
  [arXiv:1501.07274 [gr-qc]].


\bibitem{Koyama:2011xz}
  K.~Koyama, G.~Niz and G.~Tasinato,   
  Phys.\ Rev.\ Lett.\  {\bf 107} (2011) 131101 
  [arXiv:1103.4708 [hep-th]].

\bibitem{Sibandze:2016agp}
  D.~B.~Sibandze, R.~Goswami, S.~D.~Maharaj, A.~M.~Nzioki and P.~K.~S.~Dunsby,
  arXiv:1611.06043 [gr-qc];    
  D.~B.~Sibandze, R.~Goswami, S.~D.~Maharaj and P.~K.~S.~Dunsby,
  arXiv:1702.04926 [gr-qc].


\bibitem{BeltranJimenez:2017doy}
  J.~Beltran Jimenez, L.~Heisenberg, G.~J.~Olmo and D.~Rubiera-Garcia,   
  arXiv:1704.03351 [gr-qc].


\bibitem{Nzioki:2013lca}
  A.~M.~Nzioki, R.~Goswami and P.~K.~S.~Dunsby,  
  Phys.\ Rev.\ D {\bf 89} (2014) no.6,  064050 
  [arXiv:1312.6790 [gr-qc]].

\bibitem{Riotto}
  A.~Kehagias, A.~M.~Dizgah and A.~Riotto, 
  Phys.\ Rev.\ D {\bf 89} (2014) no.4,  043527 
  [arXiv:1312.1155 [hep-th]].


\end{thebibliography}
\end{document}